\documentclass[useAMS,usenatbib]{mn2e}
\usepackage{setspace}
\usepackage{graphicx}
\usepackage{caption}
\usepackage{subfigure}
\usepackage{lscape}

\begin{document}
\title[Spotted pairs with red giants in ASAS]{Orbital and physical parameters of eclipsing binaries from the ASAS catalogue -- IX. Spotted pairs with red giants \thanks{Based on observations collected through CNTAC proposals CN-2012A-21, CN-2012B-36, CN-2013A-93, CN-2013B-22, CN- 2014A-044 and CN-2014B-067, at the European Southern Observatory, Chile under programmes 089.D-0097 and 091.D-0145, and at the Subaru Telescope, which is operated by the National Astronomical Observatory of Japan, via the time exchange program between Subaru and the Gemini Observatory}}
\author[M. Ratajczak et al.] {M. Ratajczak$^{1}$\thanks{E-mail:
milena@ncac.torun.pl}, K.~G. He{\l}miniak$^{2,1}$, M. Konacki$^{1}$,  A.~M.~S. Smith$^{3}$, S.~K. Koz{\l}owski$^{1}$, \newauthor N. Espinoza$^{4}$, A. Jord\'an$^{4}$, R. Brahm$^{4}$, M. Hempel${^4}$, D.~R.~Anderson${^5}$, and C.~Hellier${^5}$\\
$^{1}$Nicolaus Copernicus Astronomical Center, Department of Astrophysics, ul. Rabia\'{n}ska 8, 87-100 Toru\'{n}, Poland\\
$^{2}$Subaru Telescope, National Astronomical Observatory of Japan, 650 North Aohoku Place, Hilo, HI 96720, USA\\
$^{3}$Nicolaus Copernicus Astronomical Center, ul. Bartycka 18, 00-716 Warsaw, Poland\\
$^{4}$Instituto de Astrof\'isica, Pontificia Universidad Cat\'olica de Chile, Av. Vicu\~na Mackenna 4860, Santiago, Chile\\
$^{5}$Astrophysics Group, Lennard-Jones Laboratories, Keele University, Keele, Staffordshire, ST5 5BG, UK}

\date{Accepted... Received...}

\pagerange{\pageref{firstpage}--\pageref{lastpage}} \pubyear{2014}

\maketitle

\label{firstpage}

\begin{abstract}

We present spectroscopic and photometric solutions for three spotted systems with red giant components. Absolute physical and orbital parameters for these double-lined detached eclipsing binary stars are presented for the first time. These were derived from the \textit{V}-, and \textit{I}-band ASAS and WASP photometry, and new radial velocities calculated from high quality optical spectra we obtained with a wide range of spectrographs and using the two-dimensional cross-correlation technique (TODCOR).
All of the investigated systems (ASAS J184949-1518.7, BQ Aqr, and V1207 Cen) show the differential evolutionary phase of their components consisting of a main sequence star or a subgiant and a red giant, and thus constitute very informative objects in terms of testing stellar evolution models. Additionally, the systems show significant chromospheric activity of both components. They can be also classified as classical RS CVn-type stars.
Besides the standard analysis of radial velocities and photometry, we applied spectral disentangling to obtain separate spectra for both components of each analysed system which allowed for a more detailed spectroscopic study. We also compared the properties of red giant stars in binaries that show spots, with those that do not, and found that the activity phenomenon is substantially suppressed for stars with Rossby number higher than $\sim$1 and radii larger than $\sim$20~R$_\odot$.
 
\end{abstract}


\section{Introduction}

The theory of stellar evolution is one of the greatest achievements of astrophysics. However, careful analysis of observations and their comparison with existing models points to some inadequacies and indicates the need to refine certain aspects, such as treatment of convection, or the issue of stellar activity. Widely-used stellar evolution models rely on accurate determinations of stellar parameters such as mass, radius, and effective temperature.  As mentioned by \citet{tor10}, only observational data yielding parameters with errors below $\sim$1-3 per cent provide sufficiently strong constraints that models with inadequate physics can be rejected.

Non-interacting eclipsing binary systems, composed of stars that have evolved as if they were single, are prime targets to retrieve physical parameters with the required accuracy. Very informative examples are systems consisting of evolved stars which have left the main sequence, e.g. red giants and subgiants. Stars belonging to this category and which have parameters known to the required precision (3 per cent or better) are valuable test beds of stellar evolution
models during evolutionary phases about which our knowledge is incomplete (approximate treatment of convection and the unsolved problem of convective core overshooting) and which is not well covered with observational data. There are just a dozen well-characterized red giants \cite[e.g.][]{and91,pie13} and just a few of them have been studied in terms of spectral analysis before. 

Systems where the components are in different phases of evolution, like AI Phe \cite[e.g.][]{and88, hel09} or ASAS-010538 \citep{rat13}, are noticeable too. The combination of a main sequence star with a red giant or subgiant is very useful for empirical verification of stellar evolution models \citep{las02}.

This paper is part of a larger effort aiming to describe an extensive sample of diverse eclipsing binaries from the ASAS catalogue \citep{hel09,hel11,rat13,hel14,hel15}. We therefore present the first detailed studies of the detached eclipsing binary systems ASAS J184949-1518.7 (hereafter ASAS-184949), BQ Aqr and V1207 Cen. First we describe our targets, then the data collection and analysis, and finally the results we obtained. Section \ref{sec_dis_gen} contains the discussion covering evolutionary status of the systems, age and distance determination, as well as the comparison of giant stars in binaries that show spots with those that do not, while Section \ref{sec_con} summarizes the conclusions.

\section{Targets}

The observing strategy includes the selection of detached eclipsing binaries (hereafter DEBs) from the extensive \textit{ASAS Catalogue of Variable Stars} \citep[ACVS;][]{poj02} and a spectroscopic campaign to infer the evolutionary status of every component and determine their physical and orbital parameters. The systems were selected on the basis of the following criteria: period P $>$ 6 days, change in brightness $<$ 1.1~mag, \textit{V}-\textit{K} $>$ 1~mag, in order to search for detached, redder systems with components of solar radius or larger. For the purposes of these studies we focused on the systems whose light curves showed out-of-eclipse time-varying brightness modulations possibly driven by the activity of the components. Thus the analysed sample includes the binary systems: ASAS-184949, BQ Aqr, and V1207 Cen.

\subsection{ASAS J184949-1518.7}
ASAS J184949-1518.7 (TYC 6861-523-1, BD-15 5108) is classified as a DEB in the ACVS. Its apparent \textit{V} magnitude is 10.29 \citep{hog00}, and the amplitude of photometric variations in \textit{V}-band is 0.22 mag. Out-of-eclipse variations in the system light curve are visible. No analysis of the system has been presented in the literature so far.

\subsection{BQ Aqr}

BQ Aqr (ASAS J233609-1628.2, TYC 6403-563-1, GSC 06403-00563, 1SWASP J233608.93-162808.3) was classified as a variable star in 1931 \citep{hof31} and its apparent \textit{V} magnitude is 10.61 \citep{hog00}, while the ACVS amplitude of photometric variations in \textit{V}-band is 0.69 mag. The target was identified as a X-ray source in ROSAT All-Sky Survey \citep{vog99} and followed up in spectroscopic survey \citep{tor06} to estimate its spectral type (K0III), although its orbital solution has not been presented in the literature yet. 

\subsection{V1207 Cen}
V1207 Cen (ASAS J142103-3253.2, TYC 7286-1252-1, 1SWASP J142101.68-325248.8) was classified as a variable by \citet{str66}. Its apparent \textit{V} magnitude is 10.69 \citep{hog00}, and the amplitude of photometric variations in \textit{V}-band is 1.02 mag. This target appears in the RAVE catalogue \citep{kor13} but no orbital solution is presented in the literature. Out-of-eclipse variations in the system light curve are noticeable.

\section{Observations}

\subsection{Photometric data}

\subsubsection{ASAS}

For the preliminary light curve (LC) analysis of the studied systems we used ASAS \textit{V}-band photometry. 570, 395, and 457 measurements were available in the ACVS for ASAS-184949, BQ Aqr, V1207 Cen, respectively. ASAS-184949 \textit{I}-band photometry \citep{sit14} yielded an additional 116 data points. The ACVS data on ASAS-184949 span more than 7 years (2001 Feb 22 to 2008 Jul 24), and almost 9 years for each of BQ Aqr (2000 Nov 21 to 2009 Jul 16) and V1207 Cen (2000 Dec 23 to 2009 Aug 13).

The LCs of all the systems show significant out-of-eclipse brightness variations related to chromospheric activity and evolving spots. For the purposes of spot-evolution studies we split the data into 4 subsets representing particular seasons based on the targets' visibility. For ASAS-184949 there are 105, 87, 86, and 161 measurements in Seasons 1 -- 4 for \textit{V}-band data, respectively. \textit{I}-band data were split into subsets defined by \textit{V}-band data seasons. The aforementioned seasons are defined as periods between 2001 Feb 22 and 2002 Dec 20 for Season 1, 2003 Feb 13 and 2004 Jul 4 for Season 2, 2004 Sep 17 and 2006 Jun 25 for Season 3, and 2005 Sep 20 and 2008 Jul 24 for Season 4. The photometric data sets for BQ~Aqr and V1207~Cen were extended by WASP measurements, which are described in Sec. \ref{phot_wasp}. ACVS data for these systems were used for a preliminary estimation of systemic period $P$ and time of minimum $T_{0}$, while the final analysis was applied just for the WASP data.

\subsubsection{WASP}
\label{phot_wasp}

Two of our targets, BQ~Aqr and V1207~Cen were also observed by the southern instrument of the Wide Angle Search for Planets (WASP; \citealt{pollacco}), WASP-South. WASP-South is located at the South African Astronomical Observatory (SAAO), near Sutherland, RSA, and consists of eight Canon 200~mm f/1.8 lenses, each equipped with a broadband filter (400 -- 700 nm), and an Andor $2048\times2048$ e2V CCD camera, on a single robotic mount. BQ~Aqr was observed a total of 20404 times between 2006 May 15 and 2009 Nov 16: 4818, 5838, 5761, and 3987 times in each of the years 2006 to 2009, which we denote Seasons 1 -- 4 respectively.

V1207~Cen was observed a total of 18811 times between 2006 May 4 and 2012 Jun 27. In each of the 2006, 2007, and 2008 seasons, this target was observed by a single camera, while in the 2011 and 2012 seasons, the object was monitored by two of the WASP-South cameras. The 2006 and 2007 seasons data were merged and denoted as Season 1 which contains 5164 measurements, while the 2008, 2011, and 2012 seasons are denoted Seasons 2 -- 4, and contain 2493, 5441, and 5713 data points, respectively.

\subsection{Spectroscopic data}

ASAS-184949, BQ Aqr, and V1207 Cen are double-lined spectroscopic binaries (SB2). In order to measure radial velocities (RVs) of the systems' components, we carried out observations using the 8.2-m Subaru telescope and the High Dispersion Spectrograph \citep[hereafter HDS; R$\sim$60\,000;][]{nog02}, the 2.2~m MPG/ESO telescope with its FEROS spectrograph \citep[R$\sim$48\,000;][]{kau99}, the 1.5-m CTIO telescope equipped with the CHIRON\footnote{Operated by the SMARTS Consortium} spectrograph \citep[service mode; R$\sim$80\,000 in slicer mode, R$\sim$25\,000 in fibre mode;][]{sch12,tok13}, and the 1.2-m Euler telescope with the CORALIE spectrograph \citep[R$\sim$60\,000;][]{que01}.

\begin{figure*}
	\begin{center}
	\begin{tabular}{cc}
	\includegraphics[scale=0.35, angle=-90]{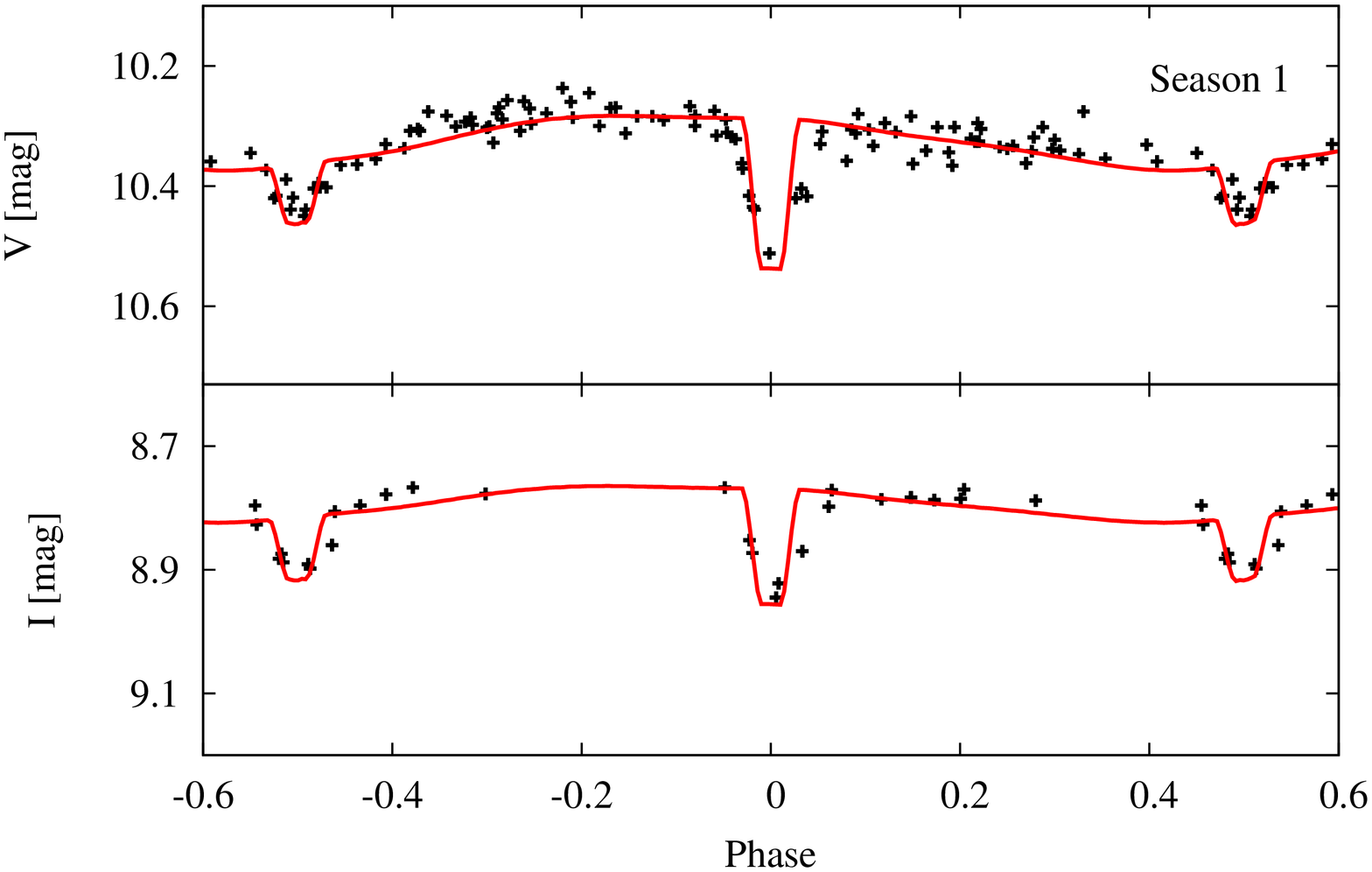}&
	\includegraphics[scale=0.35, angle=-90]{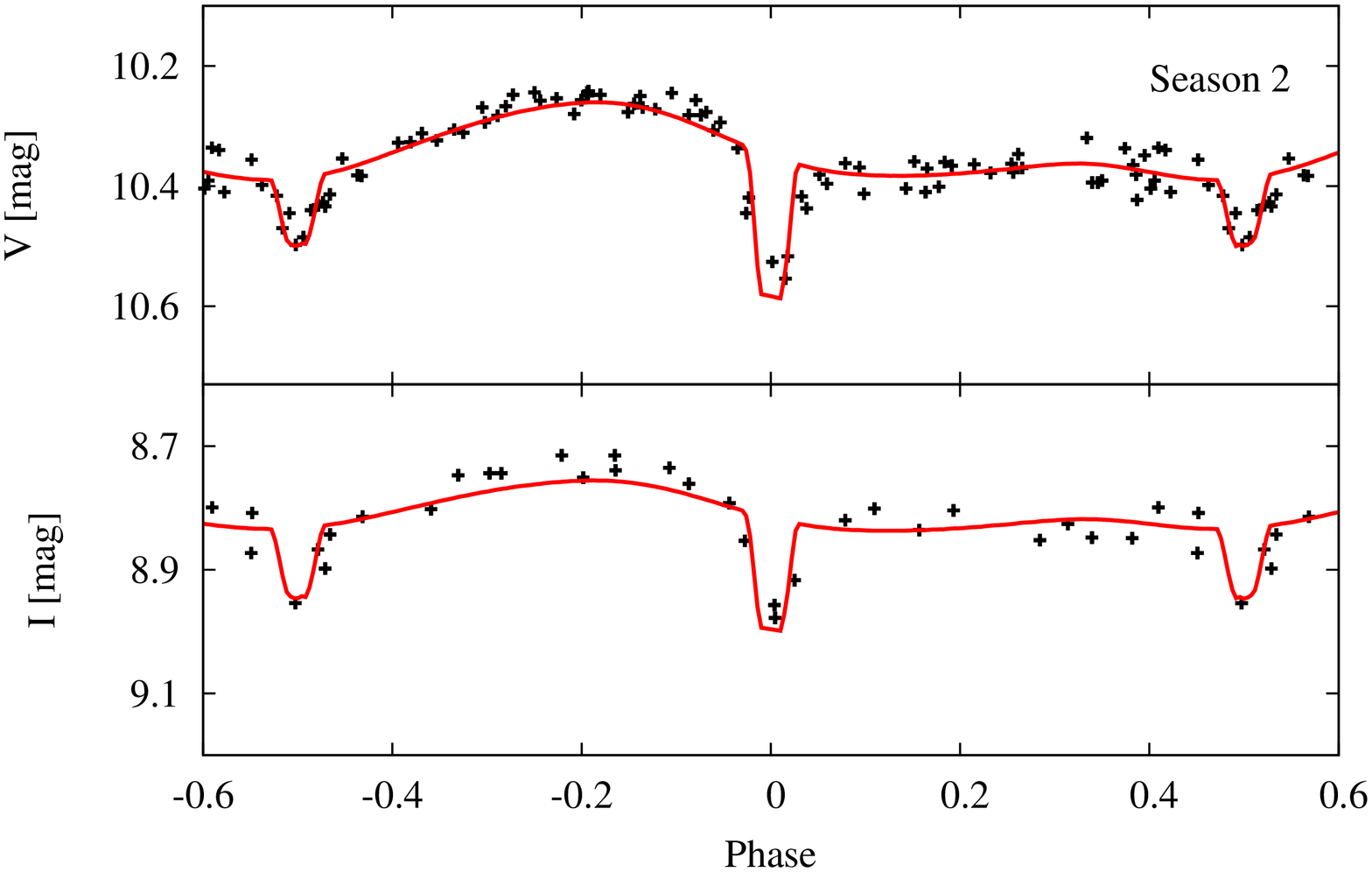}\\
	\includegraphics[scale=0.35, angle=-90]{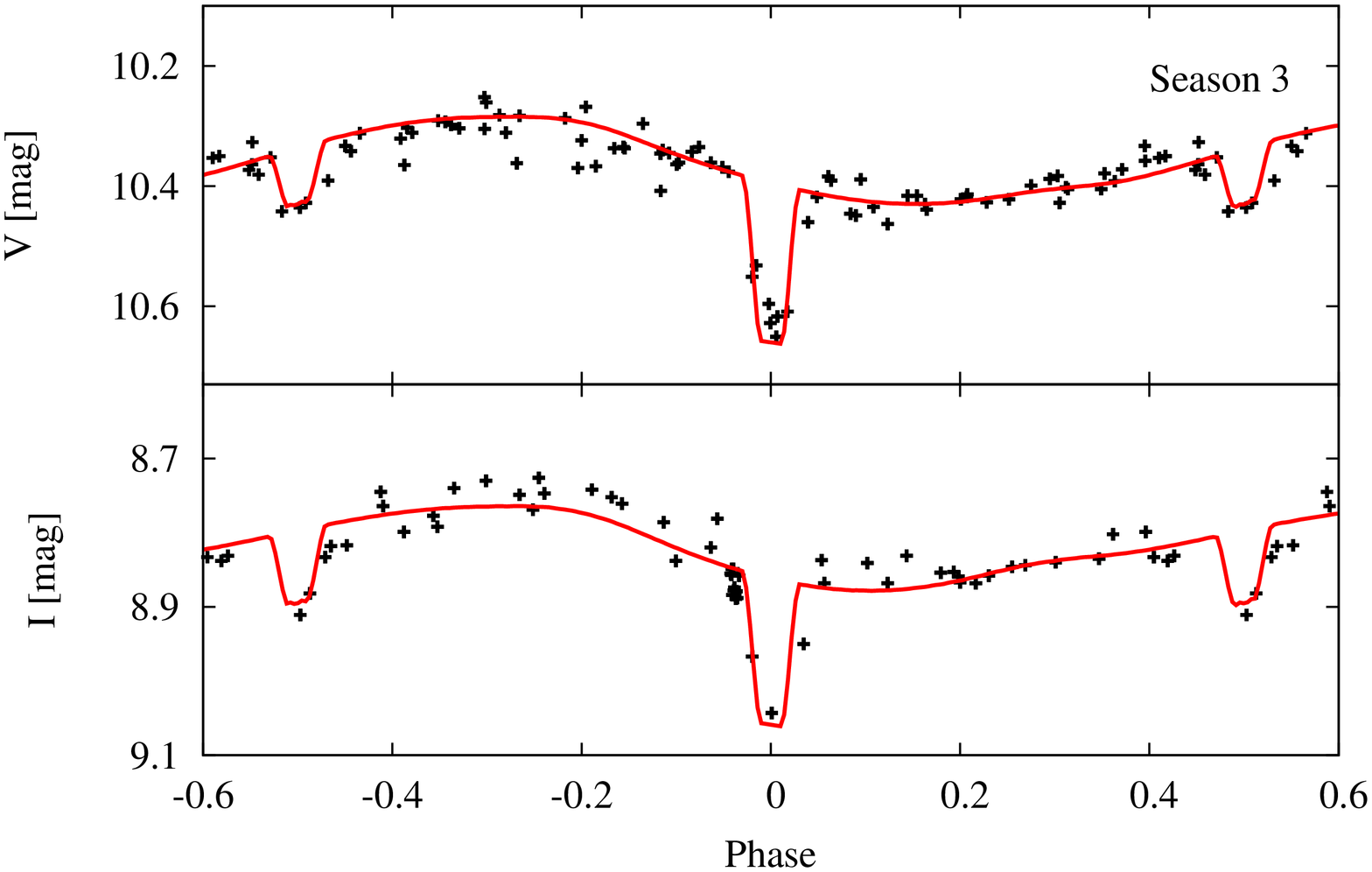}&
	\includegraphics[scale=0.35,angle=-90]{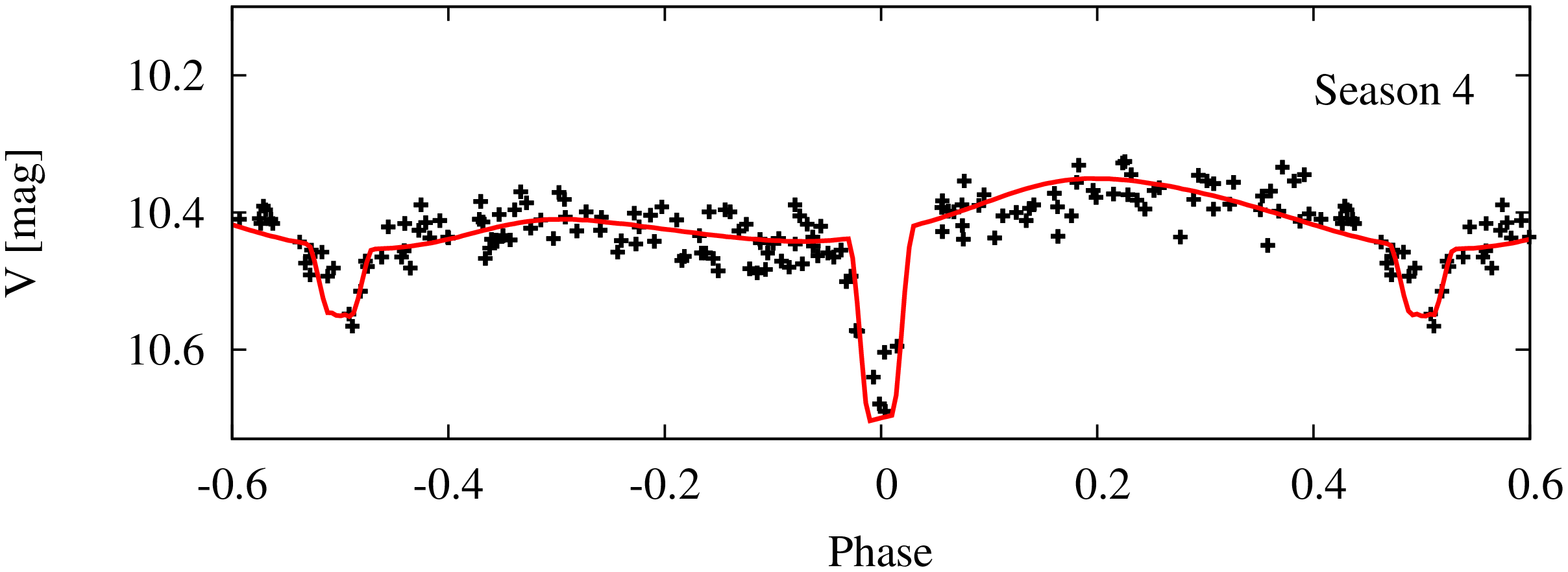}\\
	\end{tabular}
	\caption{\textit{V}- and \textit{I}-band ASAS-184949 light curves for 4 seasons with the best-fitting model. Chromospheric activity explained by the existence of evolving spots on the system components influences the light curves significantly. There is no \textit{I}-band data available for Season 4.} 
	\label{lc_184949_epochs}
	\end{center}
\end{figure*}

\begin{figure*}
	\begin{center}
	\begin{tabular}{cc}
	\includegraphics[scale=0.35, angle=-90]{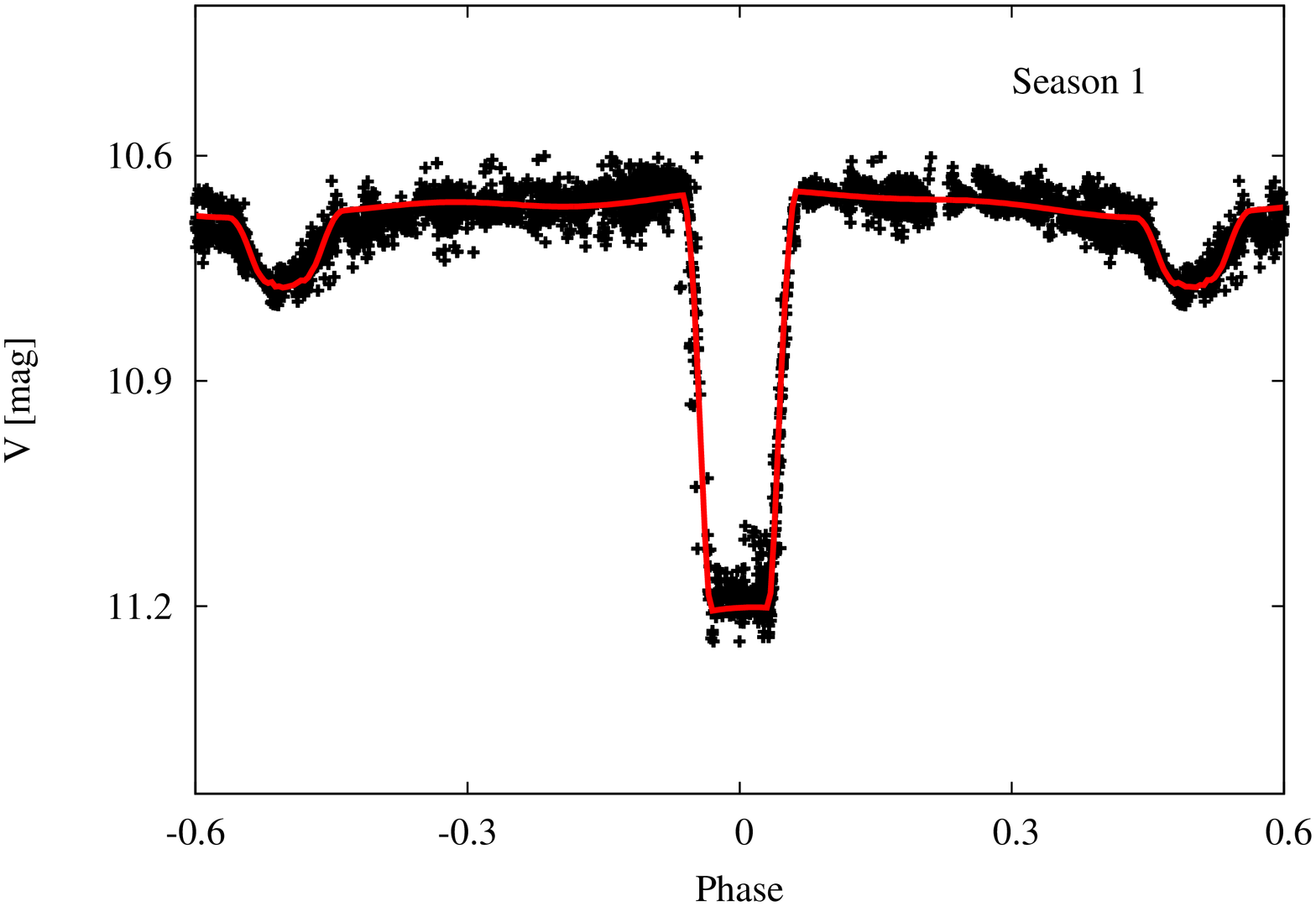}&
	\includegraphics[scale=0.35, angle=-90]{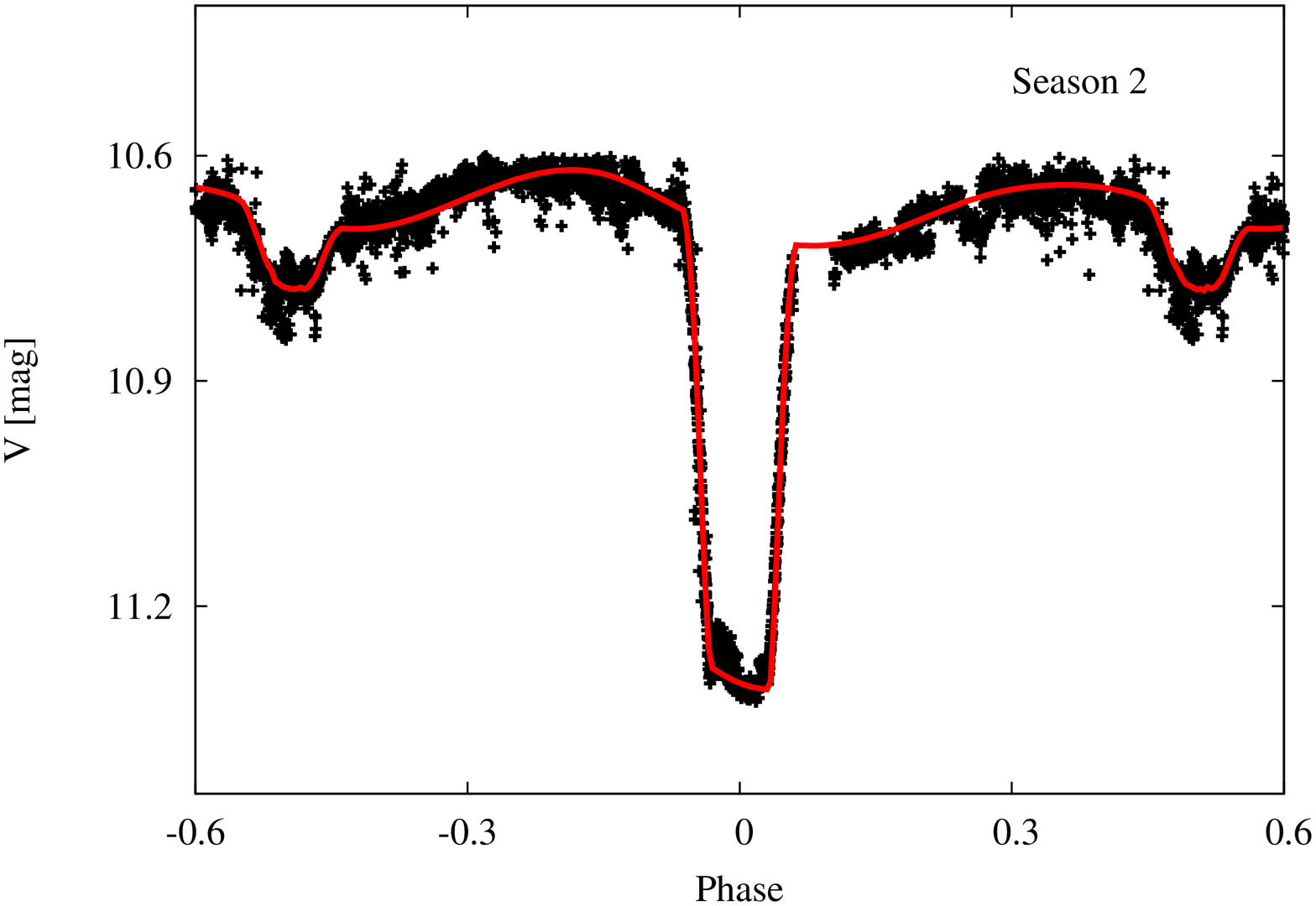}\\
	\includegraphics[scale=0.35, angle=-90]{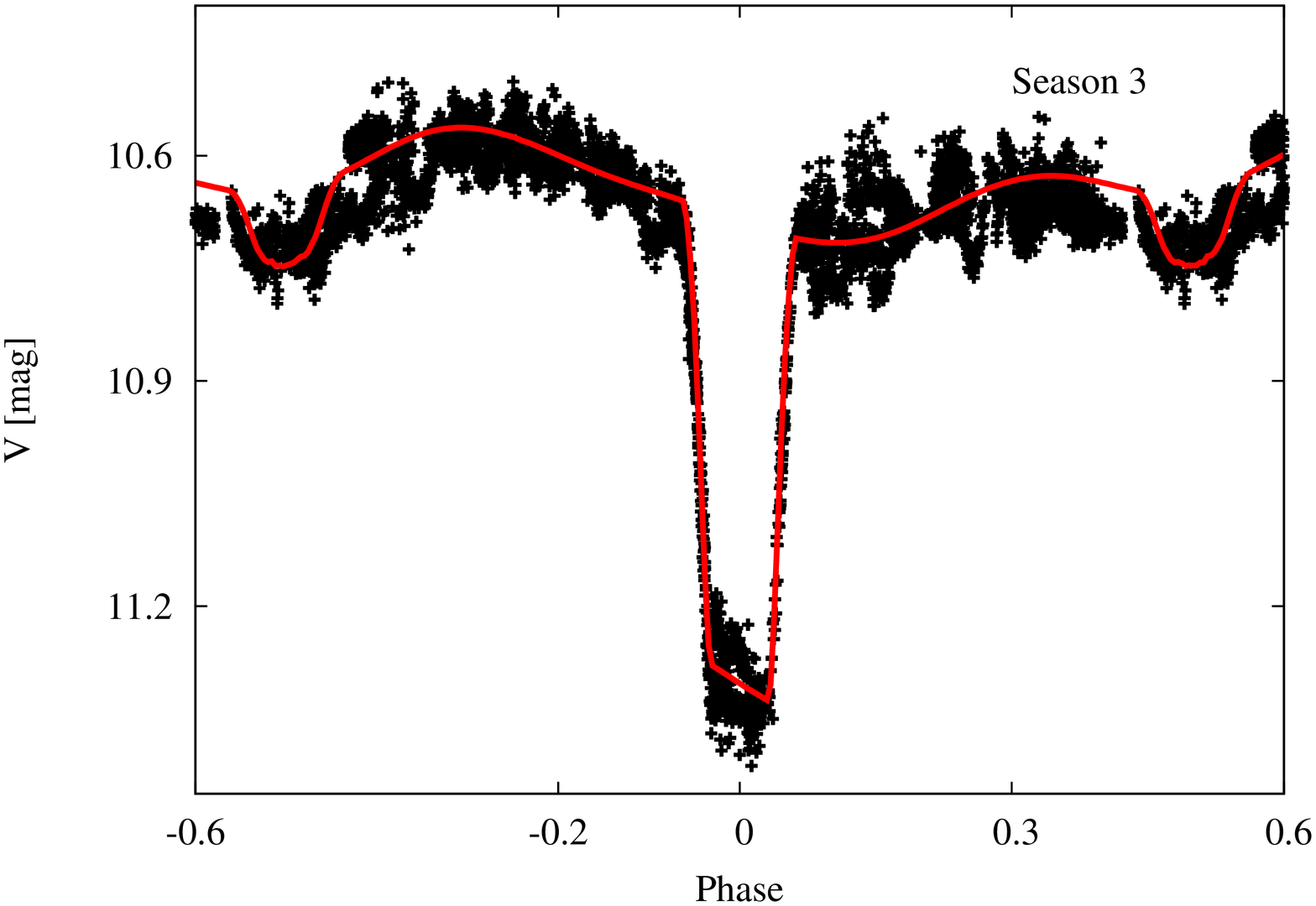}&
	\includegraphics[scale=0.35,angle=-90]{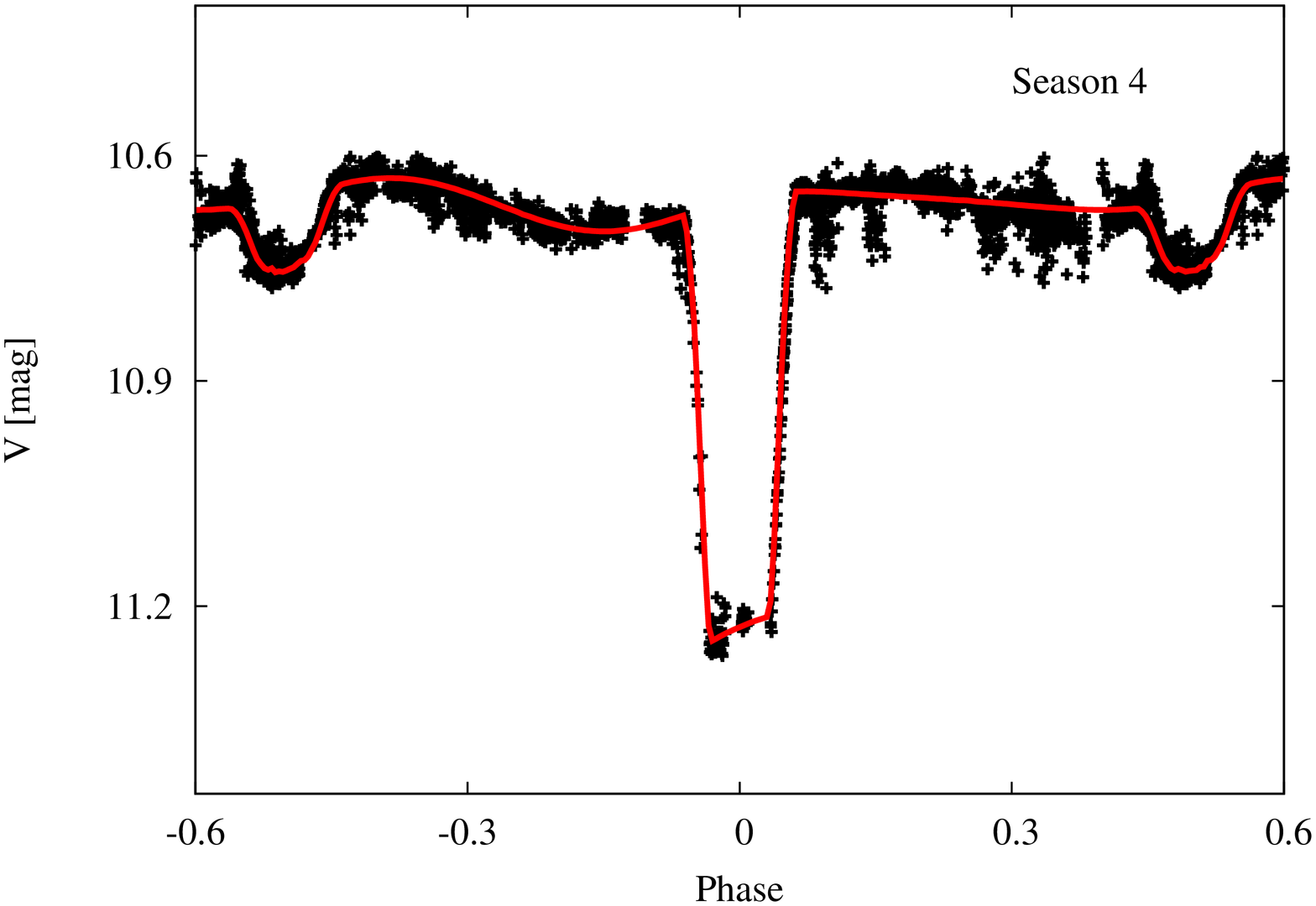}\\
	\end{tabular}
	\caption{WASP BQ Aqr light curves for 4 seasons with the best-fitting model. Chromospheric activity explained by the existence of evolving spots on the system's components influences the light curves significantly.} 
	\label{lc_233609_epochs}
	\end{center}
\end{figure*}

\begin{figure*}
	\begin{center}
	\begin{tabular}{cc}
	\includegraphics[scale=0.35, angle=-90]{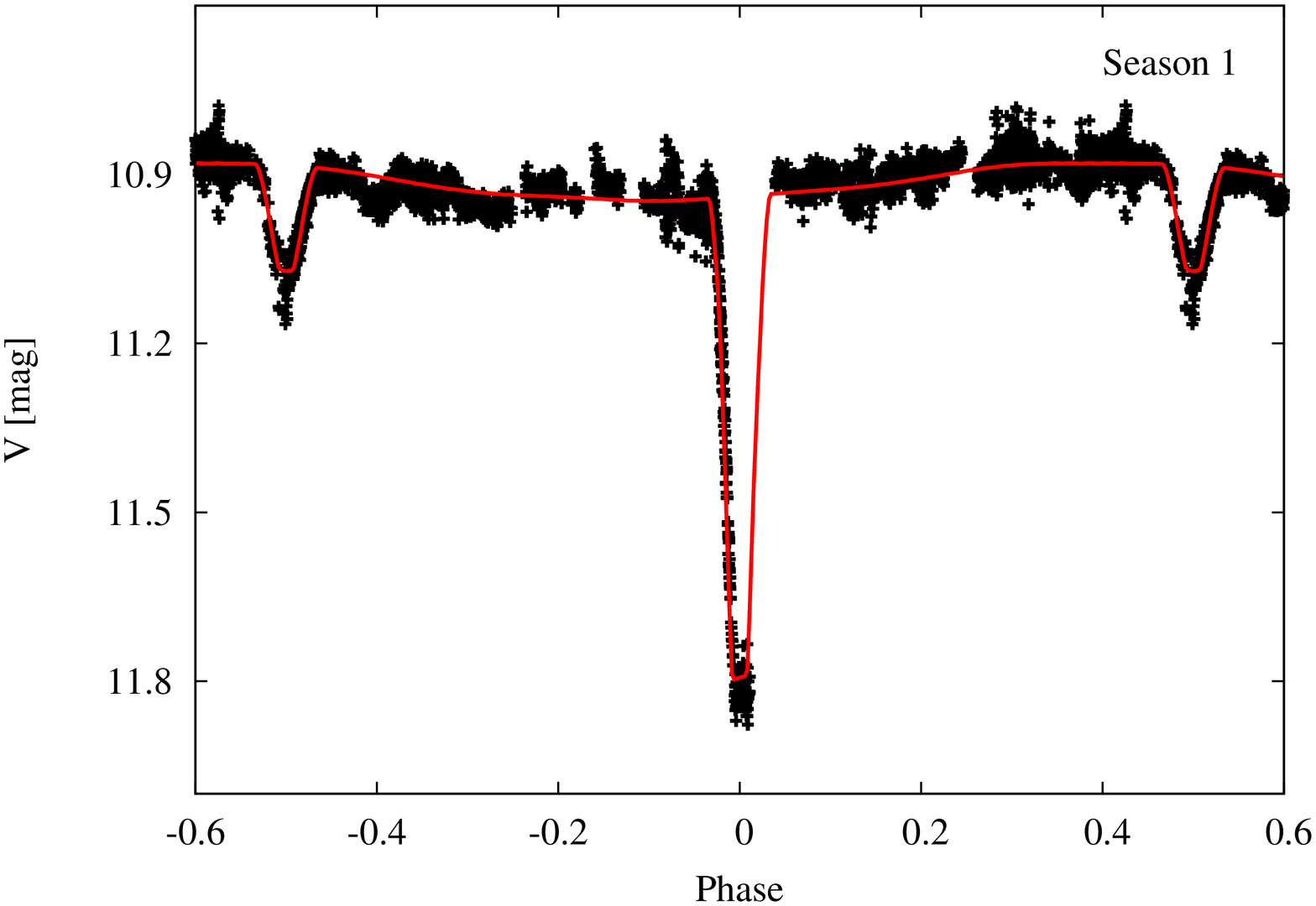}&
	\includegraphics[scale=0.35, angle=-90]{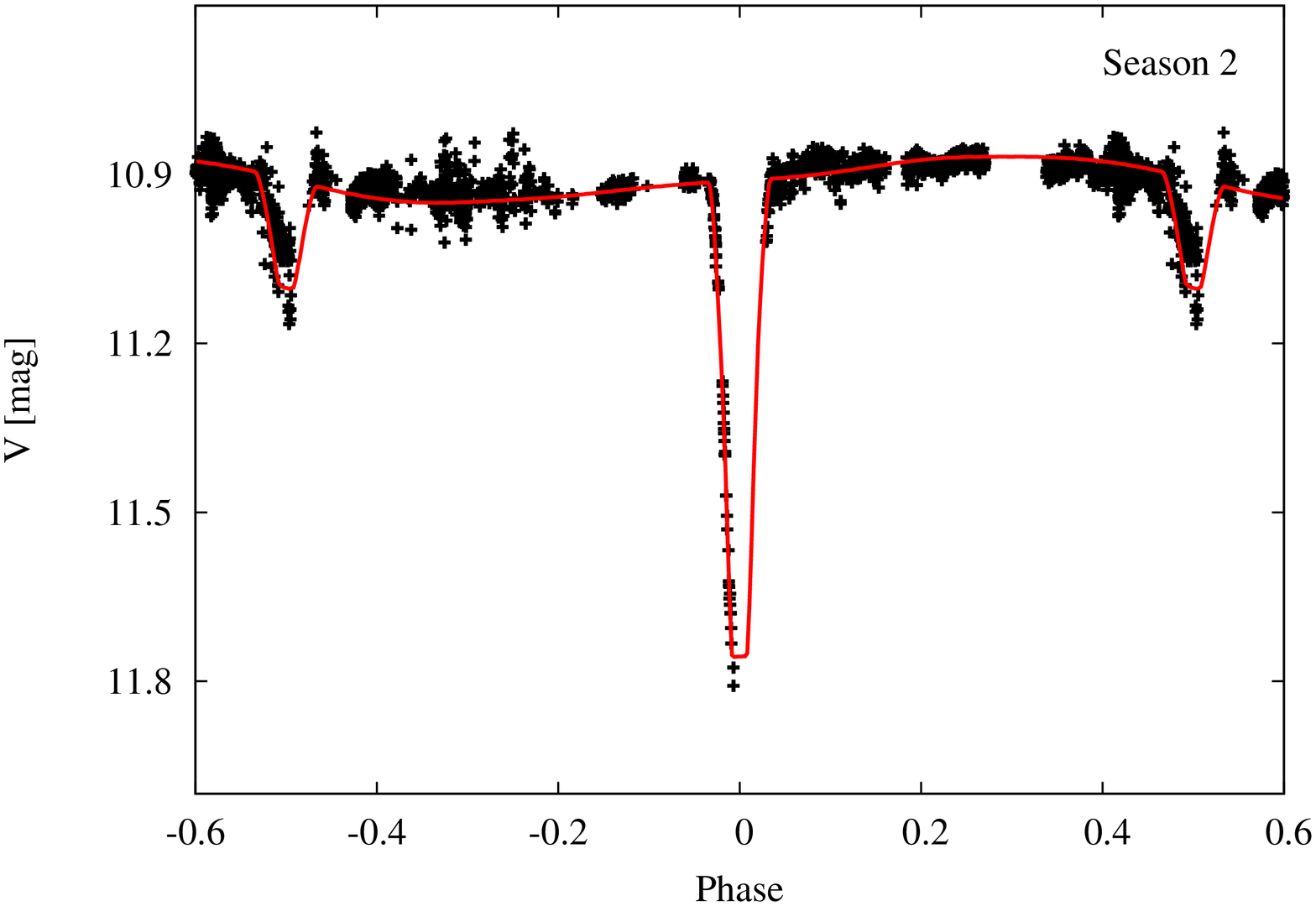}\\
	\includegraphics[scale=0.35, angle=-90]{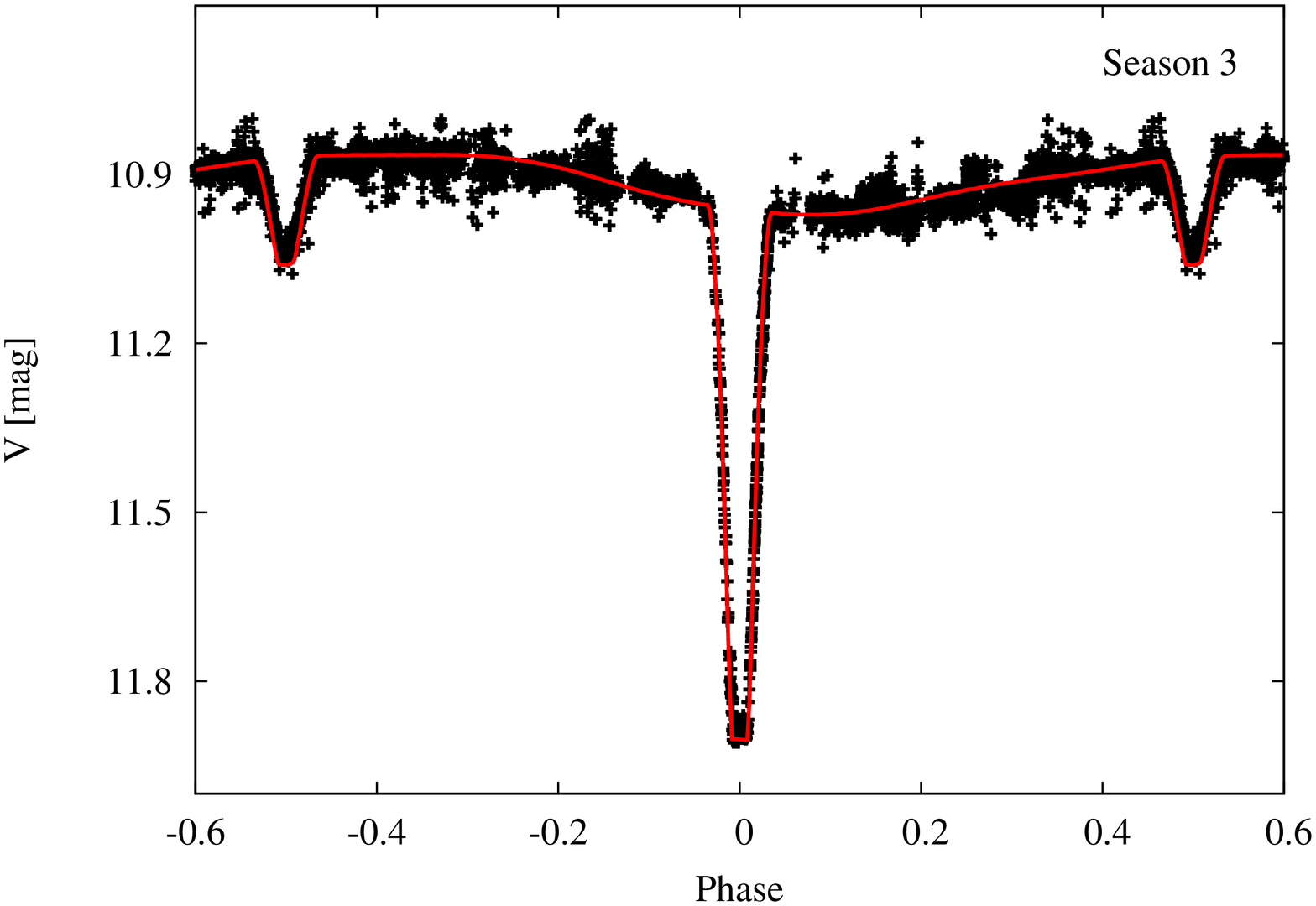}&
	\includegraphics[scale=0.35,angle=-90]{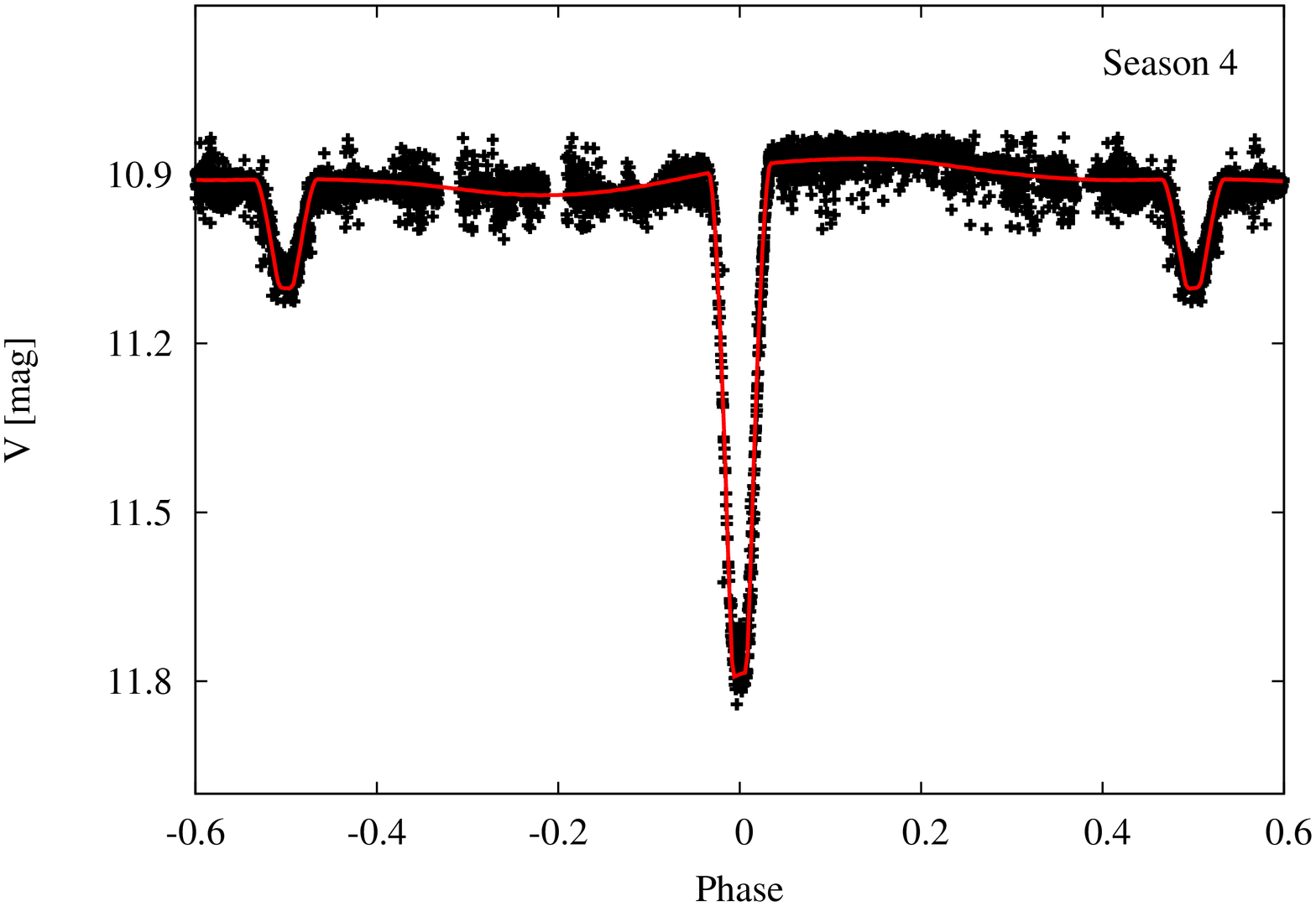}\\
	\end{tabular}
	\caption{WASP V1207 Cen light curves for 4 seasons with the best-fitting model. Chromospheric activity explained by the existence of evolving spots on the system's components influences the light curves significantly.} 
	\label{lc_142103_epochs}
	\end{center}
\end{figure*}

\subsubsection{ASAS-184949}
We obtained 31 RV measurements for each component of the ASAS-184949 system: 13 from CHIRON-fibre spectra collected in 2012-2013, eight using the CORALIE spectrograph in September 2011 and May 2012, six using the FEROS spectrograph in 2012-2013, and four from two CCD chips (blue and red) on which the spectrum taken by the HDS spectrograph at the Subaru telescope (August 2011) was recorded.

\subsubsection{BQ Aqr}
We carried out spectroscopic observations of BQ Aqr which yielded 27 RV measurements. Seventeen spectra were obtained using the CHIRON spectrograph in 2011-2013 (13 in fibre mode, four in slicer mode), and ten spectra were taken with the CORALIE spectrograph during observing runs in 2011 and 2012.

\subsubsection{V1207 Cen}
Nine RVs for V1207 Cen components come from CHIRON-fibre spectra collected in 2013 and 2014.

\section{Analysis}

\subsection{Photometry}
ASAS \textit{V}-band photometric data (fluxes and uncertainties calculated with the ASAS data reduction pipeline) were downloaded from the ACVS photometric catalogue. ASAS \textit{I}-band data were taken from ASAS 3 -- The Catalogue of Bright Variable Stars in \textit{I}- band South of Declination of {\texttt{+}}28$^{\circ}$ by \citet{sit14}, which contains measurements of brightness given with errors, obtained with the aforementioned ASAS pipeline.

The WASP data were reduced by the WASP reduction pipeline \citep{pollacco}, which produces magnitudes in the `WASP-$V$' bandpass defined by the Tycho-2 $V_t$ bandpass. Some correlated (`red') noise affecting the photometry \citep{smith} was removed by use of the {\sc SysRem} algorithm of \cite{tamuz}.

The LCs were cleaned from obvious outliers and split into subsets representing individual seasons in order to investigate the evolution of the spots. We assume that within these seasons the spots do not evolve significantly. A similar approach was used in the case of V1980 Sgr in \citet{rat13}.

\begin{figure}
\begin{center}
\begin{tabular}{c}
\includegraphics[scale=0.33,angle=270]{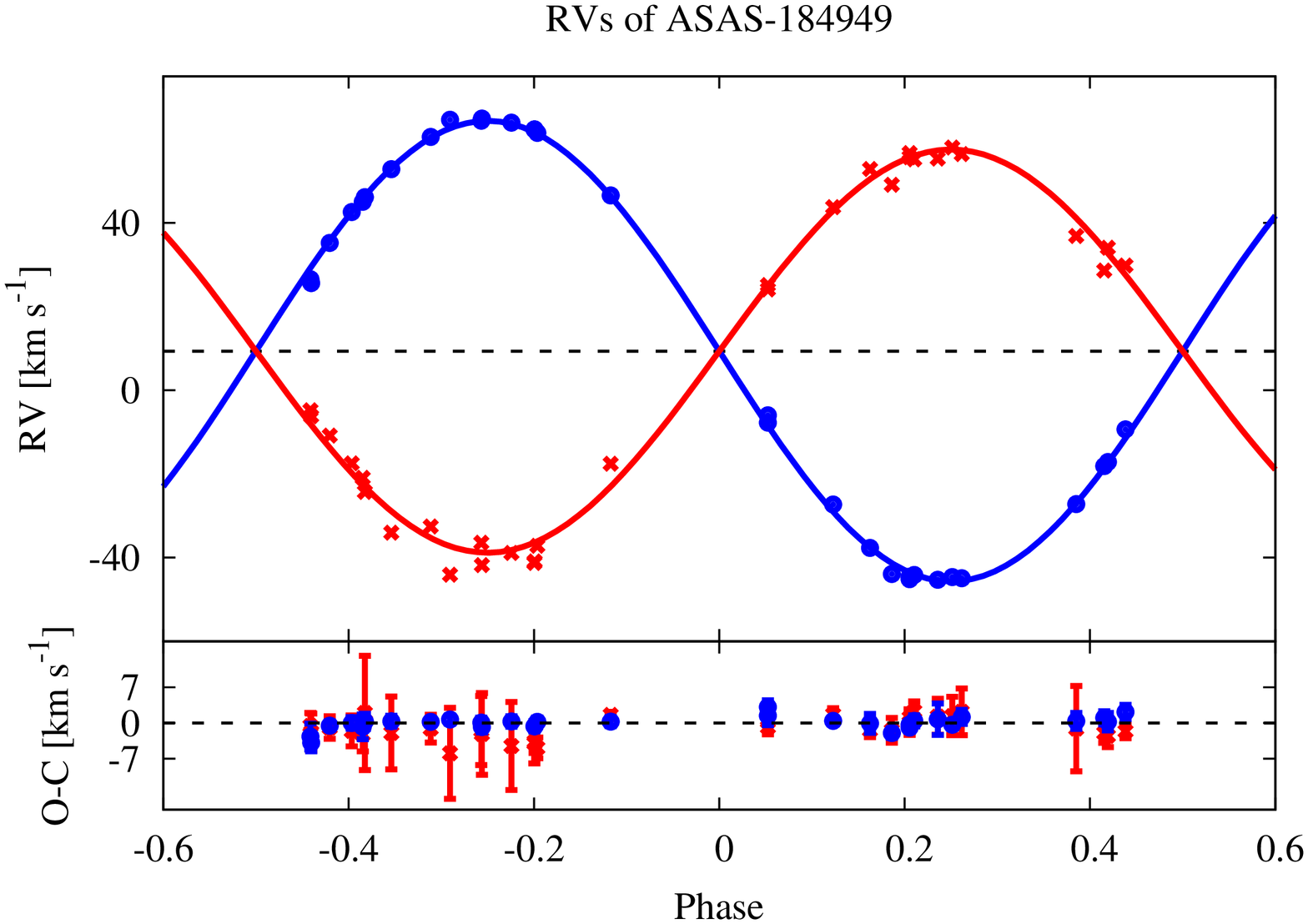}\\
\includegraphics[scale=0.33,angle=270]{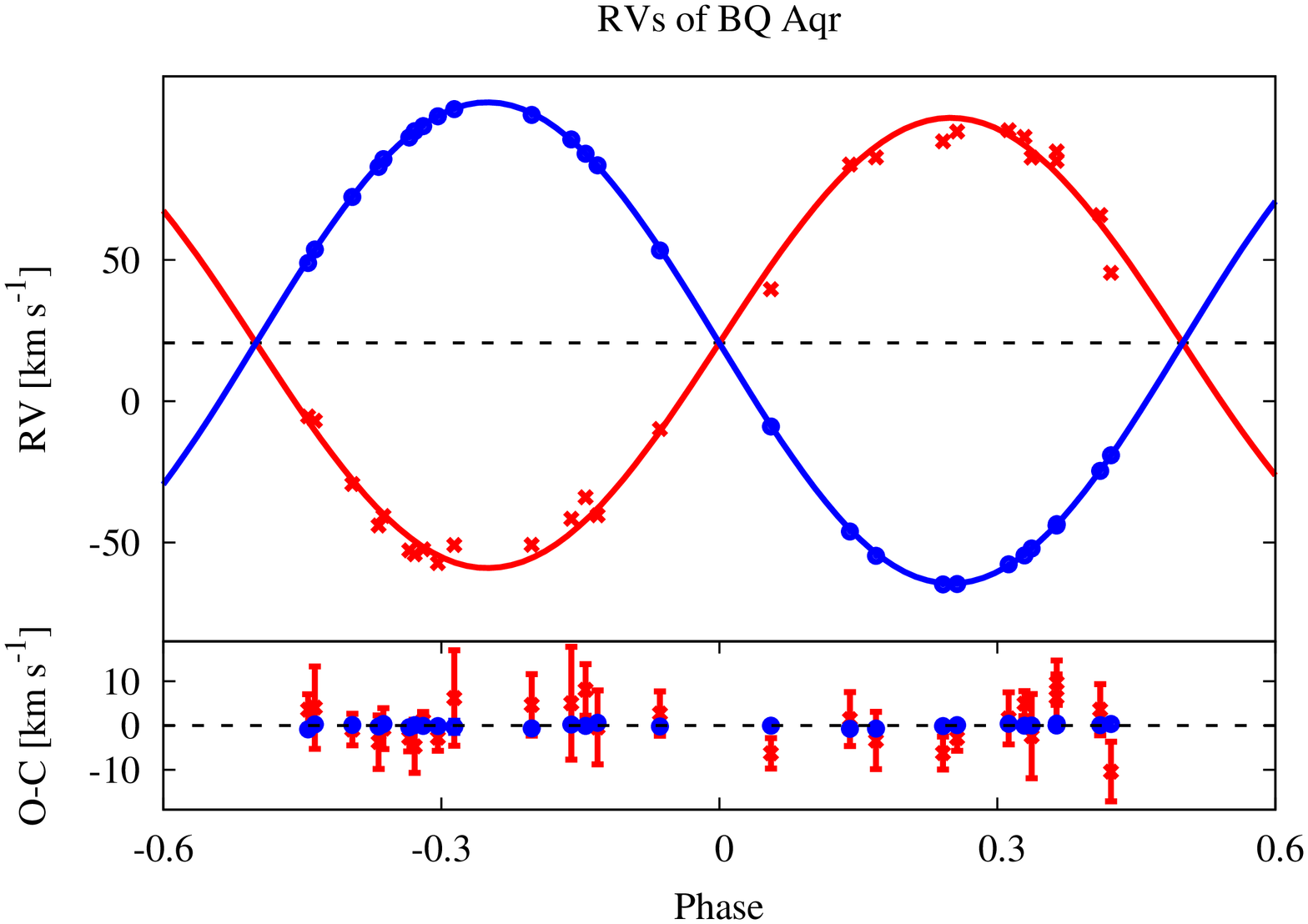}\\
\includegraphics[scale=0.33,angle=270]{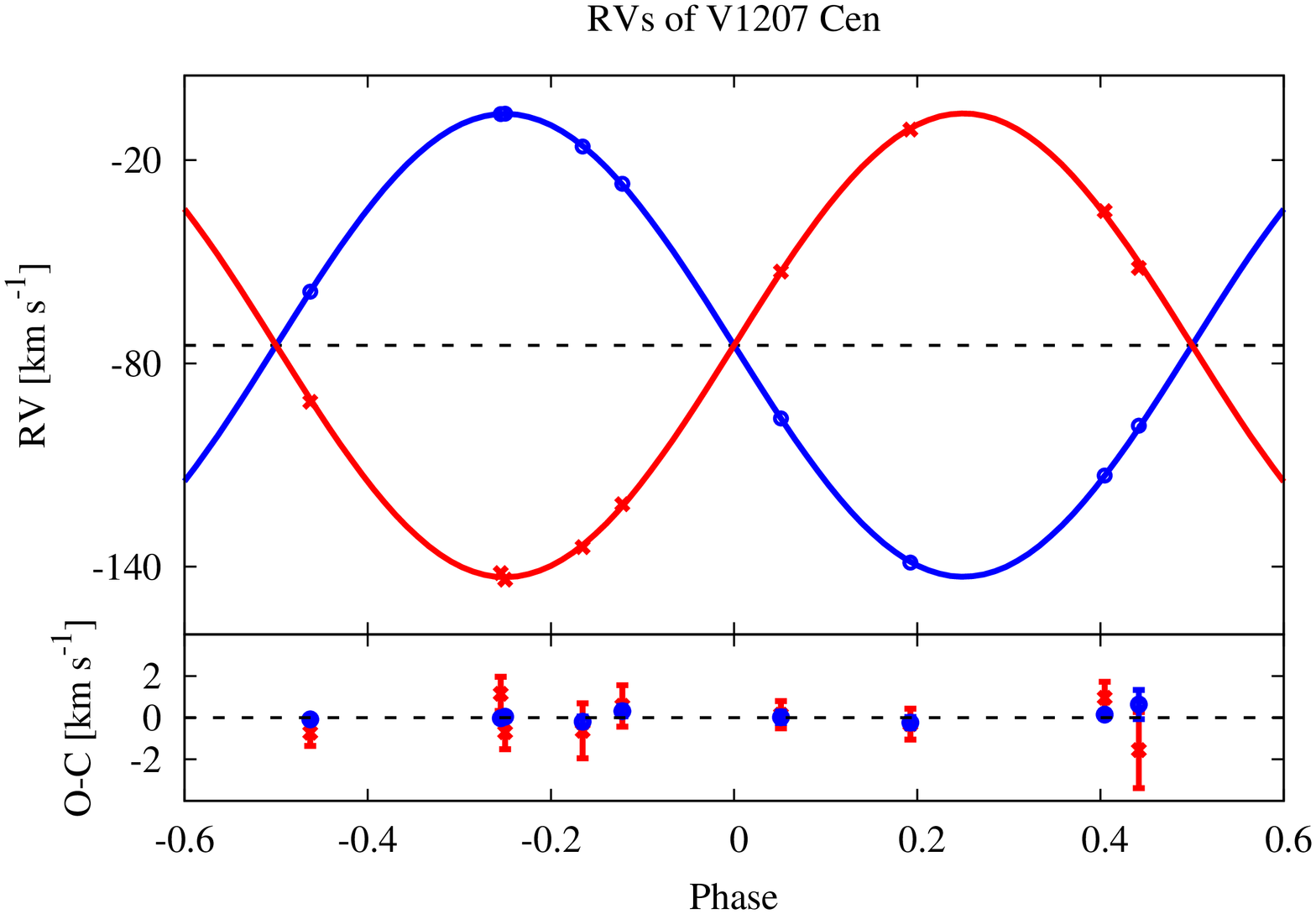}\\
\end{tabular}
\end{center}
\caption{The observed RVs of  ASAS-184949, BQ Aqr, and V1207 Cen with their best-fitting Keplerian models and O-Cs with corresponding uncertainties. Circles represent measurements of the primary and crosses measurements of the secondary.}
\label{rv_all}
\end{figure}

\subsection{Radial velocities}

\subsubsection{Data reduction}

As our work covers heterogenous spectroscopic datasets from a wide range of spectrographs, various pipelines were used to reduce and calibrate the data:

\begin{itemize}

\item{CHIRON} -- pipeline developed at Yale University \citep{tok13}. The wavelength calibration was based on the Thorium-Argon lamp exposure taken just before the science exposure with the same spectrograph settings. As the pipeline does not include barycentric velocity and time corrections, we used the \textsc{iraf}\footnote{\textsc{iraf} is written and supported by the \textsc{iraf} programming group at the National Optical Astronomy Observatories (NOAO) in Tucson, AZ. NOAO is operated by the Association of Universities for Research in  Astronomy (AURA), Inc. under cooperative agreement with the  National Science Foundation. \texttt{http://iraf.noao.edu/}} \textit{rvsao.bcvcorr} task for that purpose.  

\item{FEROS and CORALIE} --  automated pipeline developed at Pontificia Universidad Catolica de Chile, initially for CORALIE \citep{jor14}, then adopted for FEROS. It includes barycentric and continuum corrections. Spectra were taken in a simultaneous Thorium-Argon mode, where one of the fibres observes a target and the other the ThAr lamp.

\item{HDS} -- standard \textsc{iraf} procedures for echelle spectra. The wavelength calibration was done in the standard manner with Thorium-Argon lamp frames taken at the beginning and at the end of observing nights. We used the \textsc{iraf} \textit{rvsao.bcvcorr} task for barycentric velocity and time corrections.
\end{itemize}

\subsubsection{Radial velocities calculation}

Our own implementation of the two dimensional cross-correlation technique \citep[TODCOR;][]{zuc94} was applied in order to calculate RVs of the analysed stars. The method uses as references various synthetic spectra computed with the \textsc{atlas9} and \textsc{atlas12} codes \citep{kur92}. The formal RV errors were computed from the bootstrap analysis of TODCOR correlation maps created by adding randomly selected single-order maps. To obtain the best-fit with reduced $\chi^{2}$ $\approx$ 1 for our RV data and avoid error underestimation, the formal errors were multiplied by an appropriate factor. We also noted that the RV errors are much larger for the red giant components than for the main sequence/subgiant stars, mainly due to their faster rotation, likely caused by tidal locking and synchronization. 

Due to the wide range of various spectrographs we used, we also fitted an additional parameter, which allows us to compensate for different RV zero points. Initially, we set the parameter free and then shifted particular sets of data by the obtained difference in RV zero points. In the case of BQ Aqr measurements we concluded the difference in RV zero points between spectrographs is insignificant, so the final fits were done with that parameter fixed to 0.

Additionally, we found the difference between primary and secondary center-of-mass velocities ($\gamma_1$ and $\gamma_2$) is not negligible, thus we subtracted $\gamma_2$ from the secondary RVs and assumed $\gamma_1$ as a general system center-of-mass velocity. There are several possible reasons for the aforementioned shift in velocities among which template mismatch during TODCOR RVs calculations is the most significant one. Following \citet{tor09} we stress that the difference may be also caused by presence of spots on one or both components which can affect the velocities, or large scale convective motions that could be different in two stars.

\subsection{Modelling}
\label{sec_mod}

In our analysis we adopted the primary component as the star being eclipsed during the deeper (primary) eclipse (i.e. hotter) and defined $T_0$  as the time of the deeper eclipse. 

ASAS and WASP photometry was combined with the RV measurements to derive absolute orbital and physical parameters of the studied systems. The modelling procedure we used was described in detail in our previous publication -- \citet{rat13}. It applies the following codes: \textsc{v2fit} which fits a double-Keplerian RV orbit and minimizes the $\chi^{2}$ function with a Levenberg-Marquardt algorithm (RVs solution), \textsc{jktebop} \citep[v28; LC modelling, ][]{sou04a, sou04b}, \textsc{phoebe} \citep[v0.31a; \textit{Physics Of Eclipsing Binaries};][]{prs05}, and \textsc{jktabsdim} \citep[absolute values of the parameters with their uncertainties, ][]{sou04a, sou04b}. At the beginning of our analysis we found orbits consistent with circular for every system, thus we kept the eccentricity fixed at 0 further on. 
Values of rotational velocity were calculated under the assumption of tidal locking and synchronization,  which was based on the comparison of circularization and synchronization timescales (e $\sim$ 0 and $t_\mathrm{synch}$ $\textless$ $t_\mathrm{circ}$ for all systems). The tidal locking assumption is consistent with our initial \textsc{sme} (\textit{Spectroscopy Made Easy}; see further Sections) $v_\mathrm{rot}$ estimations and with the fact that wider spectral lines belong to larger components. Assuming synchronous rotation and circular orbits for eclipsing binaries ($\sin$ i $ \approx$ 1) we presume spin-orbit alignment, so $i_\mathrm{rot}$ = $i_\mathrm{orb}$, thus $v_\mathrm{rot}$ $\simeq$ $v_\mathrm{rot}$ $\sin i$.

Photometric scale factors were adjusted in \textsc{phoebe}, and gravitational darkening coefficients were set at the value of $\beta$ = 0.32 \citep{luc67}. Limb darkening was modelled using the logarithmic law of \cite{kli70}, with the values of coefficients taken from the \citet{van93} tables, while surface albedo values were assumed to be the default values from \textsc{phoebe}. The analysis showed there is no third light influence.

The mass ratio of the components obtained by applying \textsc{v2fit} was fixed in our LC modelling. We noticed a non-uniform brightness outside the eclipses, which hindered finding a consistent solution in \textsc{jktebop}, so we used this mostly for correcting the ASAS time of minimum $T_{0}$ and the period $P$. Splitting the photometric data into subsets representing seasons allowed us to study the evolution of spots on the systems' components. The LC for every season was treated as a separate set of data in \textsc{phoebe} for which the solution was found adjusting the semimajor axis \textit{a}, inclination \textit{i}, effective temperatures $T_\mathrm{eff}$, luminosity levels, and gravitational potentials $\Omega$, not including stellar spots in a model. The systemic velocity $\gamma_1$ and mass ratio were fixed in \textsc{phoebe} modelling with the values obtained with \textsc{v2fit}. Then one season (hereafter base season) was chosen on the basis of the largest data phase coverage and the least-significant spots influence, for which the spots analysis was carried out (for ASAS-184949 and BQ Aqr that was Season 1, for V1207 Cen -- Season 4). We kept the same number of spots on a given component as for the base season in other seasons, but we adjusted the spots' radii, temperatures and locations, to find stellar parameters for each season. Final values of the systems' \textit{a}, \textit{i}, and potentials were adopted as weighted means of these quantities from different seasons. The first estimation of the brighter component effective temperature was based on the colour-temperature calibration \citep{wor11} using TYCHO-2 colours \citep{hog00}, but for the final analysis we used component temperatures obtained by performing spectral analysis using \textsc{sme} (described in Sec. \ref{sp_an}). \textit{I}-band data were fitted simultaneously with \textit{V}-band datasets during the \textsc{phoebe} modelling.

The solutions obtained for the photometric measurements are presented in Fig. \ref{lc_184949_epochs}, \ref{lc_233609_epochs}, and \ref{lc_142103_epochs}. RV curves for ASAS-184949, BQ Aqr, and V1207 Cen are presented in Fig. \ref{rv_all}. The mean formal errors of photometric  and RVs  measurements, \textsc{RMS} of orbital fitting, and multiplicative factors for analysed systems are presented in Table \ref{tab_fit}.

\subsubsection {Spectral analysis}
\label{sp_an}

For all systems, the phase coverage of the spectroscopic observations from the same spectrograph (CHIRON), taken in the same mode (here: fibre mode) was sufficient to disentangle the components spectra. The spectral disentangling technique \citep{bag91, kon10} was used to extract the individual contributions of both stars to the composite spectra and reconstruct the spectra of each component. It has been proven \citep{ili04} that in the case of time-independent component fluxes, spectral disentangling can be performed assuming equal fluxes of the components, and the resulting spectra can be renormalized afterwards. The two separate spectra obtained in this way were scaled by their brightness ratio (hereafter $BR$, defined as primary over secondary brightness ratio) and used for individual studies of every star. Spectral analysis was performed leading to a description of the stellar atmospheres. As significant brightness modulations due to the spots made estimation of $BR$ of the systems components troublesome, we adopted the values taken from the TODCOR analysis ($BR_\mathrm{A184949}$ = 0.70, $BR_\mathrm{BQAqr}$ = 0.67, $BR_\mathrm{V1207Cen}$ = 0.72). For detailed spectral analysis we used the software package Spectroscopy Made Easy \citep[hereafter \textsc{sme},][]{val96, val98}.

Following the claim presented in the work of \citet{val05} that line blending becomes more severe in the blue and in cooler stars, making continuum placement and derived parameters less accurate, we analysed only 7 rows (over 38) of the spectra covering the wavelength region from 5\,927 to 6\,399 \AA.  We used the list obtained from the Vienna Atomic Line Database \citep[VALD,][]{pis95, kup99} as the atomic line data with the initial values described by \citet{val05} and \citet{kur92} model atmospheres.

The spectral synthesis yielded effective temperatures and metallicities for both components of the analysed systems. Keeping the values of log $g$ that we determined in the first stage of the \textsc{phoebe} analysis, rotational velocities calculated under the assumption of tidal locking (which were consistent with preliminar spectral analysis results), and solar abundances pattern fixed \citep{gre07}, we fitted for $T_\mathrm{eff}$ and $[M/H]$. As starting values, we used the temperatures we adopted from \textsc{phoebe} analysis \citep[brighter component effective temperature based on the colour-temperature calibration from ][]{wor11} and solar metallicity. Calculations were made for each of 7 spectral orders, before finally averaging the results. The variance between different orders was taken as the uncertainty of every parameter.

\begin{table*}
\caption{Mean formal errors of photometric ($\bar{\sigma}_{\mathrm{LC}}$) and RVs ($\bar{\sigma}_{\mathrm{RV}}$) measurements, \textsc{RMS} of orbital fitting (\textsc{RMS}$_\mathrm{RV}$), and multiplicative factors ($\mathrm{MF_{RV}}$) for analysed systems. A-V stands for ASAS-\textit{V} band, A-I - ASAS-\textit{I}, and W - WASP data.}
\label{tab_fit}
\begin{tabular}{cccc}
\hline
\hline
{Parameter} & {ASAS-184949} & {BQ Aqr} & {V1207 Cen}\\
\hline
$\bar{\sigma}_{\mathrm{LC}} [\mathrm{mag}]$ &  0.03 (A-V), 0.07 (A-I)  & 0.16 (A-V), 0.07 (W) & 0.15 (A-V), 0.15 (W) \\
$\bar{\sigma}{_\mathrm{RV1}} [\mathrm{km~s^{-1}}]$ & 0.3 & 0.3  & 0.1 \\
$\textsc{rms}{_\mathrm{RV1}} [\mathrm{km~s^{-1}}]$ &  1.3  &  0.4 & 0.3 \\
$\mathrm{MF_{RV1}}$ &  4.0  & 1.7 & 1.9 \\
$\bar{\sigma}{_\mathrm{RV2}} [\mathrm{km~s^{-1}}]$ &  0.9 & 1.3 & 0.5\\
$\textsc{rms}{_\mathrm{RV2}} [\mathrm{km~s^{-1}}]$ &  2.4  & 4.9  & 0.9 \\
$\mathrm{MF_{RV2}}$ &  4.0  & 4.6 & 1.9 \\

\hline
\end{tabular}
\end{table*}

\begin{table*}
\caption{Orbital and physical parameters of ASAS-184949, BQ Aqr, and V1207 Cen. Values of effective temperatures $T_\mathrm{eff}$ and metallicities $[M/H]$ are taken from spectral analyses of disentangled spectra. The values after the slash of $T_\mathrm{eff}$, $[M/H]$, $L$, and $M_\mathrm{bol}$ for the primary component of ASAS-184949 and BQ Aqr, and both components of V1207 Cen describe alternative solutions for the systems (see Sec. \ref{sec_dis}).}
\label{tab_orb}
\begin{tabular}{lrlrlrl}
\hline
\hline
Parameter & \multicolumn{2}{c}{ASAS-184949} & \multicolumn{2}{c}{BQ Aqr} & \multicolumn{2}{c}{V1207 Cen}\\
\hline
$P$ [d] &  35.7091 & $\pm$0.0011 & 6.6205062 & $\pm$0.0000021 & 8.5365711 & $\pm$0.0000011\\
$T_{0}$ [JD-2450000]	& 2063.214  &$\pm$0.022 & 1885.443 & $\pm$0.007 & 1906. 0056 & $\pm$0.0025\\
$K_{1}$ [km~s$^{-1}$]  & 55.02 &  $\pm$0.27 & 85.18 & $\pm$0.09 & 68.5 & $\pm$0.4 \\
$K_{2}$ [km~s$^{-1}$] & 48.2 & $\pm$0.6 & 79.7 & $\pm$1.1 & 68.33 & $\pm$0.11 \\
$\gamma_{1}$ [km~s$^{-1}$]& 9.3 & $\pm$0.5 & 20.6 &$\pm$1.8 & -74.65 &$\pm$0.09\\
$\gamma_{2}$ [km~s$^{-1}$]& -3.8 & $\pm$0.6 & -1.5 &$\pm$1.9 & 1.15 &$\pm$0.29\\
$e$ 		 	&  0.0  & fixed & 0.0 & fixed & 0.0 & fixed\\
$i$  & 85.7 & $\pm$1.9 & 89.5 & $\pm$0.5 & 88.6 & $\pm$0.5\\
$a$ [R$_\odot$] & 72.69 & $\pm$0.19 & 21.66 & $\pm$0.08 & 23.08 & $\pm$0.05 \\
$M_{1}$ [M$_\odot$] & 1.91 & $\pm$0.05 & 1.49 & $\pm$0.04 & 1.132 & $\pm$0.008\\
$M_{2}$ [M$_\odot$] & 2.19 & $\pm$0.04 & 1.588 & $\pm$0.021 & 1.134 & $\pm$0.013\\
$R_{1}$ [R$_\odot$]  & 3.0 & $\pm$0.7 & 2.072 & $\pm$0.014 & 1.92 & $\pm$0.06\\
$R_{2}$ [R$_\odot$]  & 9.0 & $\pm$3.0 & 6.53 & $\pm$0.28 & 3.003 & $\pm$0.026 \\
log $g_{1}$ [cm~s$^{-1}$]  & 3.76 & $\pm$0.19 & 3.977 & $\pm$0.006 & 3.924 & $\pm$0.026\\
log $g_{2}$ [cm~s$^{-1}$]  & 2.8 & $\pm$0.4 & 3.01 & $\pm$0.04 & 3.537 & $\pm$0.011\\
$v_\mathrm{rot1}$ [km~s$^{-1}$] & 4.3 & $\pm$0.9 & 15.83 & $\pm$0.11 & 11.4 & $\pm$0.3\\
$v_\mathrm{rot2}$ [km~s$^{-1}$] & 13.0 & $\pm$5.0 & 49.9 & $\pm$2.6 & 17.79 & $\pm$0.21\\
$T_\mathrm{eff1}$ [K] & 5560/6630 & $\pm$400/320 & 6260/6390 & $\pm$240/230 & 5780/6340 & $\pm$230\\
$T_\mathrm{eff2}$ [K] & 4560 & $\pm$120 & 4490 & $\pm$230 & 4560/5000 &$\pm$160\\
$[M/H]_{1}$ & -0.41/0.35 & $\pm$0.26/0.22 (fixed)& 0.46/0.12 & $\pm$0.22/0.11 (fixed) & -0.16/-0.45 &$\pm$0.15/0.11 (fixed)\\
$[M/H]_{2}$ & 0.35 & $\pm$0.22 & 0.12 & $\pm$0.11 & -0.45 &$\pm$0.11\\
log $L_{1}$ [L$_\odot$] & 0.89/1.20 & $\pm$0.23/0.21 & 0.77/0.80 & $\pm$0.06/0.06 & 0.51/0.73 &$\pm$0.06/0.07\\
log $L_{2}$ [L$_\odot$] & 1.54 & $\pm$0.38 & 1.19 & $\pm$0.09 & 0.57/0.71 &$\pm$0.06/0.06\\
$M_\mathrm{bol1}$ [mag] & 2.5/1.7& $\pm$0.6/0.5& 2.81/2.72 &$\pm$0.16/0.15 & 3.47/2.93 &$\pm$0.14/0.16\\
$M_\mathrm{bol2}$ [mag] & 0.9 & $\pm$0.9 & 1.76 & $\pm$0.23 & 3.34/2.98 &$\pm$0.14/0.15\\
$R_{1}/R_\mathrm{Roche}$  & 0.11 &  & 0.26 & & 0.21 &\\
$R_{2}/R_\mathrm{Roche}$  & 0.33 &  & 0.79 & & 0.34 &\\
\hline
\end{tabular}
\end{table*}

\section{Results}
\label{res}

The physical and orbital parameters of ASAS-184949, BQ Aqr, and V1207 Cen with their 1$\sigma$ uncertainties are presented in Table \ref{tab_orb}. Radii are given in R$_\odot$ units, as well as fractions of Roche limit (Roche radius, R$_\mathrm{Roche}$). The Roche radius is defined as 
$R_\mathrm{Roche}/a = 0.49 q^{2/3} / [0.6 q^{2/3}+\ln(1+q^{1/3})]$ \citep{egg83} where $a$ is the semi-major axis and $q$ is the mass ratio. Effective temperatures and metallicities are based on the spectral analysis we performed using \textsc{sme}. In order to find distances to the investigated systems we estimated reddening using maps of dust IR emission by \citet{sch98} recalibrated by \citet{sch11} and then used bolometric corrections by \citet{bes98} to convert bolometric magnitudes into absolute visual magnitudes.

\subsection{ASAS-184949}

The components' masses are 1.91 M$_\odot$ and 2.19 M$_\odot$, while the radii are 3.0 R$_\odot$ and  9.0 R$_\odot$ for the primary and secondary, respectively. The uncertainties in masses reach 3 per cent and 2 per cent. Unfortunately the quality of the photometric data (just ASAS photometry) and LC modulation makes the determination of radii imprecise, with uncertainties of 22 per cent and 40 per cent.

\begin{figure*}
\begin{tabular}{cc}
\includegraphics[angle=-90,width=0.5\textwidth,clip=true]{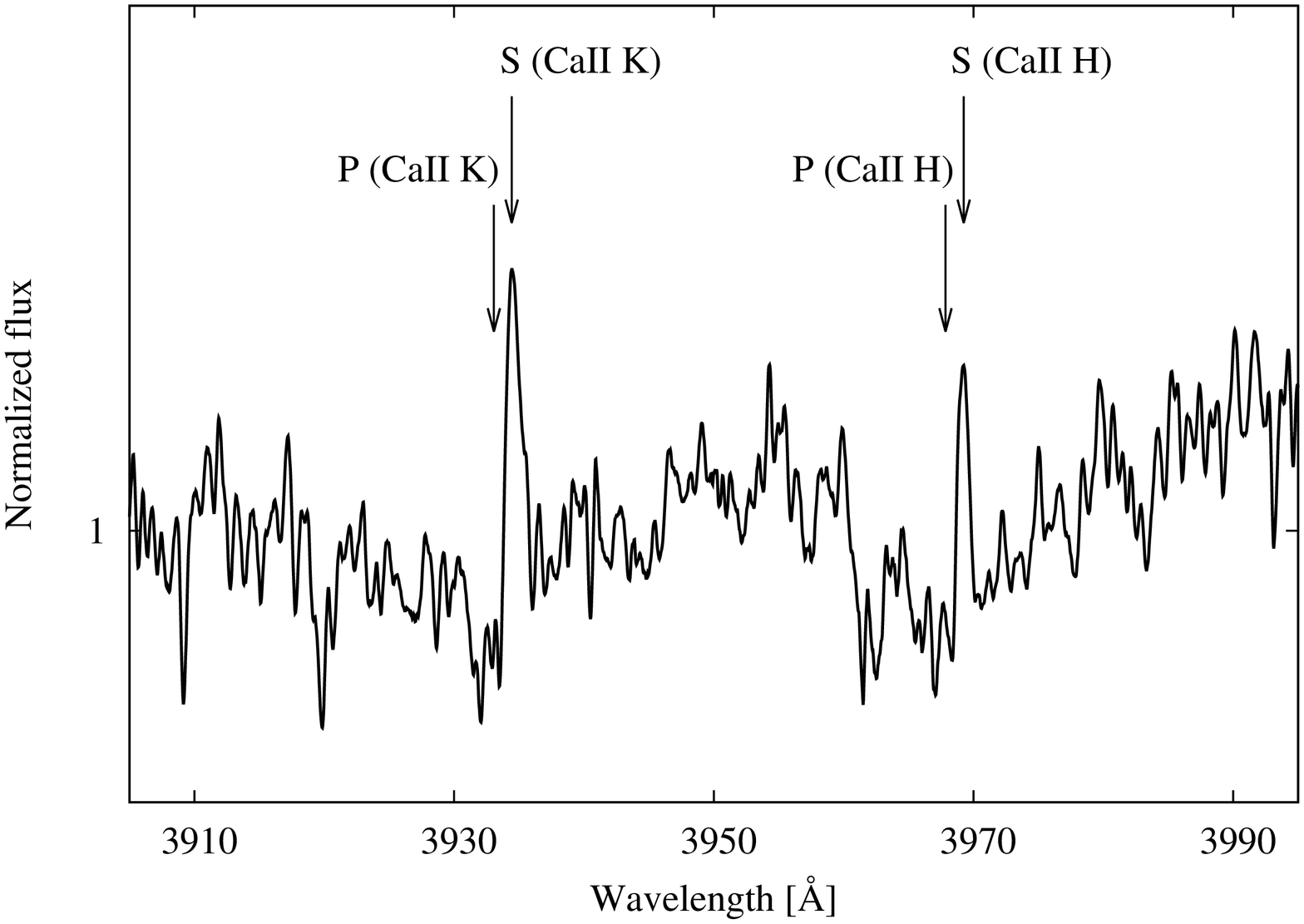}
\includegraphics[angle=-90,width=0.5\textwidth,clip=true]{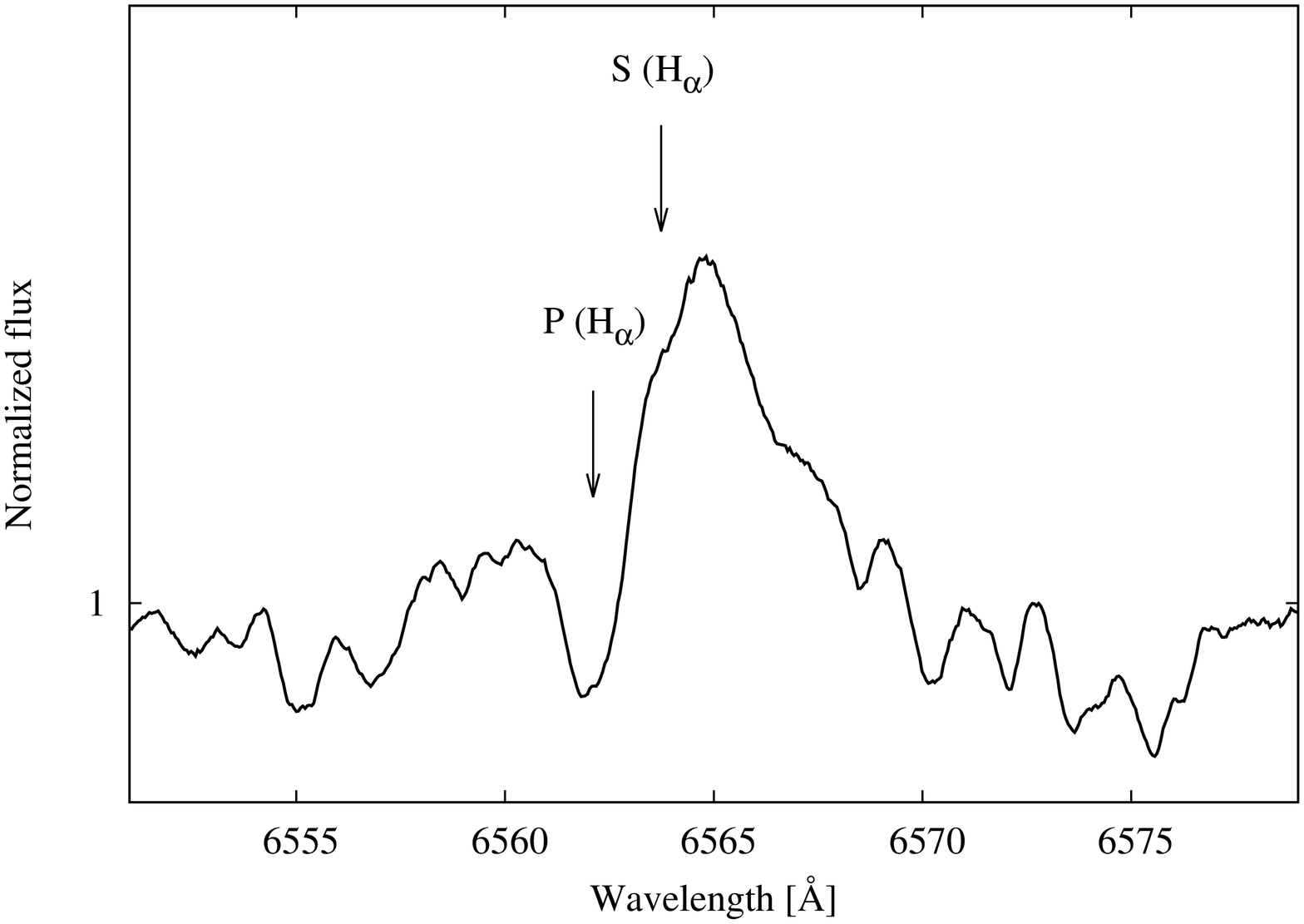}
\end{tabular}
\caption{Regions of Ca~\textsc{II} H and K (left panel), and $\mathrm{H_{\alpha}}$ (right panel) of continuum normalized FEROS spectra of ASAS-184949. Primary star features are labelled with P, and secondary star features with S, the arrows correspond to RVs of each component inferred from the orbital solution. Emission in secondary star features is significant. The left panel represents a spectrum from 2013 May 16 ($\mathrm{RV_{1}}$ = -49.2 km~s$^{-1}$ , $\mathrm{RV_{2}}$ = 56.5 km~s$^{-1}$), and the right panel from 2012 Jun 24 ($\mathrm{RV_{1}}$ = -31.8 km~s$^{-1}$, $\mathrm{RV_{2}}$ = 42.2 km~s$^{-1}$). Spectra were smoothed with a 10-pixel boxcar.} 
\label{ca_emission}
\end{figure*}

Strong variability of $\mathrm{H_{\alpha}}$ (6\,563~\AA) line and emission in its spectral region was detected for ASAS-184949 in HDS spectra. Additionally we noticed emission in Ca~\textsc{II} K (3\,934~\AA), and Ca~\textsc{II} H (3\,969~\AA) lines in FEROS spectra (Fig. \ref{ca_emission}). Such features can be indicators of an active chromosphere for late-type stars \citep{str94}, and thus the presence of spots on the stellar surfaces which cause significant brightness modulations. The \textsc{phoebe} spots analysis resulted in a model with two spots -- one located on a primary, the second on the secondary component.

Effective temperatures and metallicities were determined by applying \textsc{sme} to the disentangled spectra and yielded $T_{1}$ = 5560 $\pm$400 K, $T_{2}$ = 4560 $\pm$ 120 K, $[M/H]_{1}$ = -0.41 $\pm$ 0.26, and $[M/H]_{2}$ = 0.35 $\pm$ 0.26 (hereafter solution A). The inconsistency we found in the metallicity of the two components led us to an alternative solution (resulting with $T_{1}$ = 6630 $\pm$ 320 K and system $[M/H]$ of 0.35) which is described in Sec. \ref{sec_dis} (hereafter solution B).  The distances we derived are $d_\mathrm{1849494}$ = 496 $\pm$ 129 and 508 $\pm$ 133 pc for the solutions A and B, respectively.

The resulting \textsc{rms} of the residuals to the radial velocity fits are 1.3 km~s$^{-1}$  and 2.4 km~s$^{-1}$ for the primary and secondary components, respectively. The average photometric error is 0.03 mag (\textit{V}-band) and 0.07 (\textit{I}-band).

\subsection{BQ Aqr}
BQ Aqr components' masses are 1.49 M$_\odot$ and 1.59 M$_\odot$, while the radii are 2.07 R$_\odot$ and 6.53 R$_\odot$ for the primary and secondary, respectively. The masses are determined with a precision of 1 and 3 per cent and uncertainties in the radii reach 1 per cent for the primary and 4 per cent for the secondary.

As the Ca~\textsc{II}~K and H region is beyond the CORALIE and CHIRON spectral ranges, we looked at the Balmer lines, but no clear emission was detected. Although no evident traces of chromospheric activity were found in any spectra, the existence of spots on the secondary component was noticeable in the shape of a cross correlation function. Thus we applied spots to the LC analysis and found a model with three spots (one on the primary, two on the secondary) as the most accurate one.

Effective temperatures determined from the spectral analysis are 6260 $\pm$ 240 K for the primary, and 4490 $\pm$ 230 K for the secondary, and the metallicities we derived are of $[M/H]_{1}$ = 0.46 $\pm$ 0.22, and $[M/H]_{2}$ = 0.12 $\pm$ 0.11. For further analysis we decided to assume the system metallicity as the secondary metallicity value, $[M/H]$ = 0.12, because evolutionary tracks calculated for both the primary's metallicity, and the mean of the metallicity values obtained in \textsc{sme} for both components, do not have fitted parameters determined for both components.  Adopting $[M/H]$ = 0.12, we performed \textsc{sme} spectral analysis obtaining a new value of effective temperature for the primary $T_1$ = 6390 $\pm$ 230 K. The derived distance to the system is $d_\mathrm{BQ Aqr}$ = 617 $\pm$ 75 pc.

The resulting \textsc{rms} in the residuals of the fits to the radial velocities are 0.4 km~s$^{-1}$  and 4.9 km~s$^{-1}$ for the primary and secondary components, respectively. The average photometric error is 0.16 mag (ASAS) and 0.07 (WASP). 

\subsection{V1207 Cen}
The component masses we obtained are almost equal of 1.13 M$_\odot$ with 1 per cent uncertainties. The radii of the system's components are 1.92 R$_\odot$ for the primary, and 3.00 R$_\odot$ for the secondary. The uncertainties reach 1--3 per cent. 

We noted strong variability in the secondary $\mathrm{H_{\alpha}}$ line, including emission in some of the spectra. Other activity indicators, like emission in the Ca \textsc{II} K and H region are beyond CHIRON's spectral range. \textsc{phoebe} was used to find the best fitting model with spots for the system (one big spot on the secondary, and a small one on the primary).

Effective temperatures and metallicities determined from disentangled spectra using \textsc{sme} are $T_{1}$ = 5780 $\pm$ 230 K, $T_{2}$ = 4560 $\pm$ 160 K, $[M/H]_{1}$ = -0.16 $\pm$ 0.15, and $[M/H]_{2}$ = -0.45 $\pm$ 0.11. Evolutionary tracks calculated either using the primary value or the mean of metallicities showed poor agreement to fitted parameters for both components. The secondary value of $[M/H]$ = -0.45 was therefore chosen for further analysis. We found the effective temperature estimation to be in disagreement with the models, so we adopted an alternative solution with $T_{1}$ = 6340 K and $T_{2}$ = 5000 K  (see Sec. \ref{sec_dis}) as the final one. We stress that, despite the components having nearly equal masses, the temperatures must be very different, because the LC eclipses show very unequal depths.

The resulting \textsc{rms} in the residuals of the fits to the radial velocities are 0.3 km~s$^{-1}$  and 0.9 km~s$^{-1}$ for the primary and secondary components, respectively. The average photometric error is 0.15 mag (for both ASAS and WASP data). The derived distance is of $d_\mathrm{V1207 Cen}$ = 447 $\pm$ 31 pc.

\begin{figure*}
\begin{tabular}{cc}
\includegraphics[angle=-90,width=0.5\textwidth,clip=true]{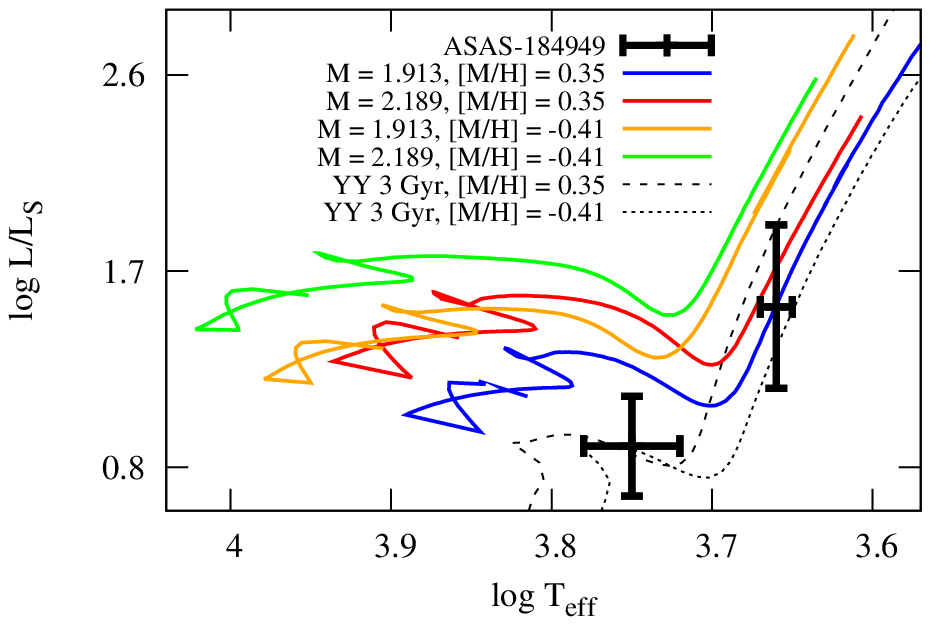}
\includegraphics[angle=-90,width=0.5\textwidth,clip=true]{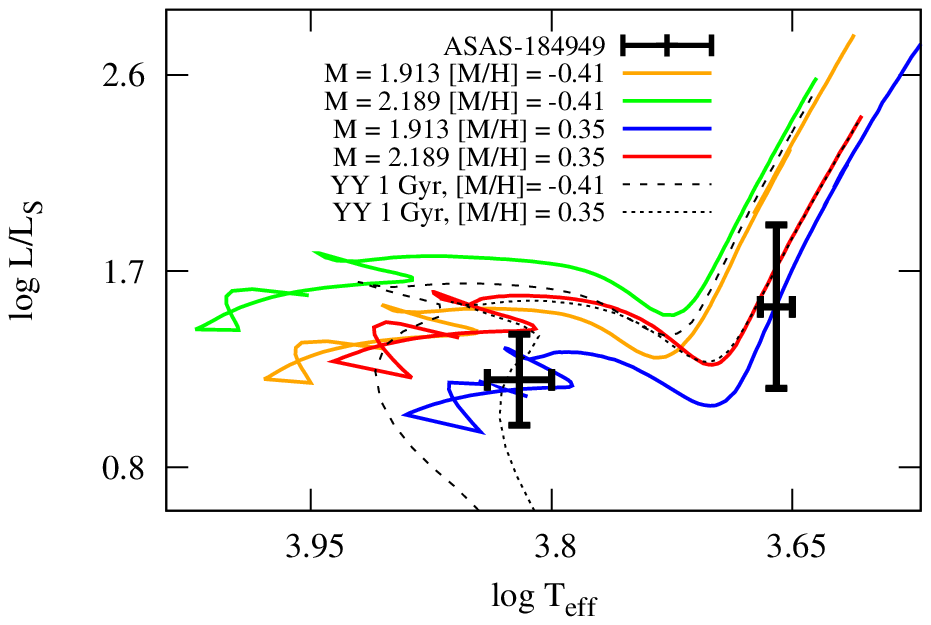}
\end{tabular}
\caption{Two parameters solutions (left panel: solution A -- $T_\mathrm{eff1}$ = 5560; right panel: solution B -- $T_\mathrm{eff2}$ = 6630 K) and YY evolutionary tracks for ASAS-184949 for $[M/H]$ of -0.41 (green and orange) and 0.35 (blue and red). Dashed lines represent YY isochrones calculated for a given metallicity.} 
\label{184949_track}
\end{figure*}

\begin{figure*}
\begin{tabular}{cc}
\includegraphics[angle=-90,width=0.5\textwidth,clip=true]{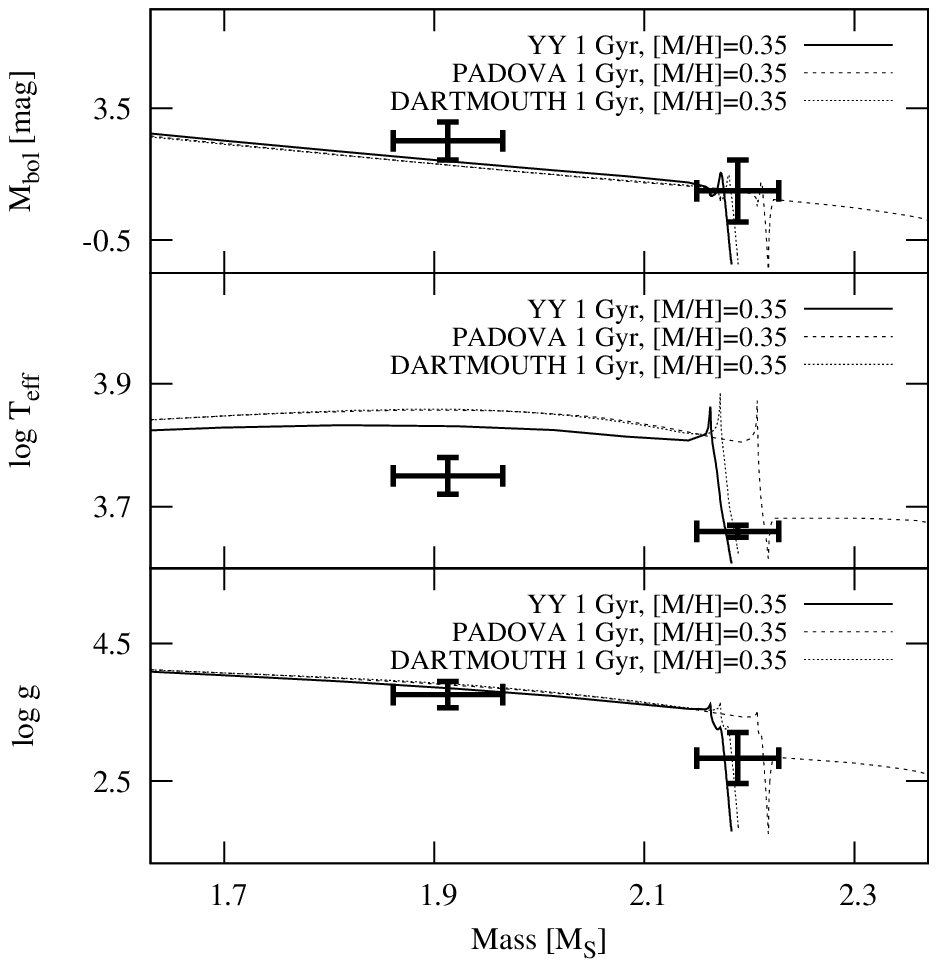}
\includegraphics[angle=-90,width=0.5\textwidth,clip=true]{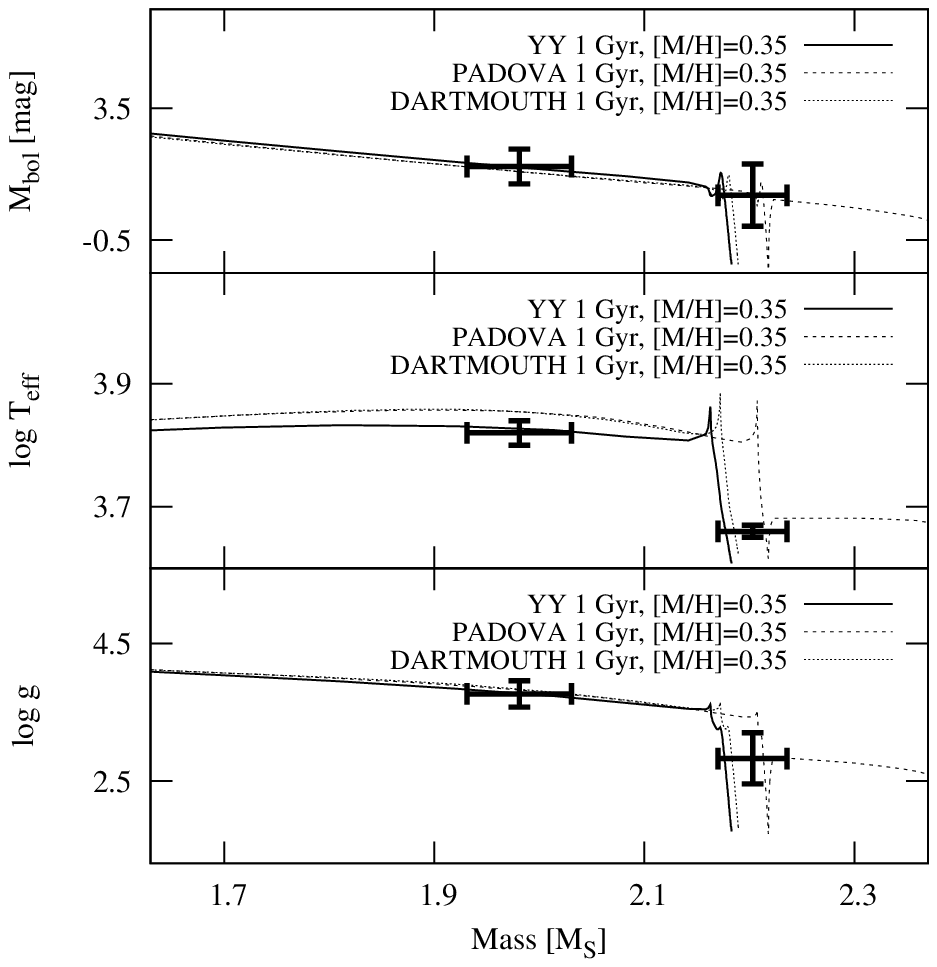}
\end{tabular}
\caption{ASAS-184949 components parameters (left panel: solution A; right panel: solution B) with the best fitting YY, PARSEC, and Dartmouth isochrones of age of 1 Gyr, and metallicity $[M/H]$ = 0.35.} 
\label{184949_iso_kompil}
\end{figure*}

\section{Discussion}
\label{sec_dis_gen}

\subsection{Age and evolutionary status}

\label{sec_dis}

To check the evolutionary status of the stars the Yonsei-Yale \citep[hereafter YY, ][]{yi01} evolutionary tracks for $\alpha$-enhancement of 0 were applied. As there is no mass-loss on the red giant branch assumed in YY models, we estimated the mass-loss using a new formulation of Reimers' Law from \citet{sch05} and found it to be negligible, at least up to the point specifically reached by the giants under scrutiny, thus the application of YY tracks is justified. 

Assuming that the system components are coeval and have the same metallicity, we faced the problem of inconsistent metallicity determination for ASAS-184949 ($[M/H]_{1}$ = -0.41, $[M/H]_{2}$ = 0.35). To test which value is more reliable, tracks for both values were interpolated. Tracks calculated for $[M/H]$ = -0.41 (primary metallicity obtained applying \textsc{sme}) do not fit the derived parameters (hereafter solution A), unlike the more reliable tracks obtained by applying $[M/H]$ = 0.35 (see left panel of Fig. \ref{184949_track}), thus we assumed this value as the proper one. As a next step, we performed an \textsc{sme} analysis keeping the metallicity value $[M/H]$ = 0.35 fixed and obtained new values of $T_\mathrm{eff}$ for the primary. An alternative solution (hereafter solution B) based on the newly-computed value (new values of $T_\mathrm{eff1}$, $L_\mathrm{1}$, $M_\mathrm{bol1}$ with errors) is described in Tab. \ref{tab_orb} after the slash. Both solutions with corresponding evolutionary tracks and isochrones are presented in Fig. \ref{184949_track}. The resulted evolutionary tracks for the more credible solution B indicate that the primary component is leaving the main sequence or is in a subgiant phase, while the secondary is already on the red giant branch. 

The system age estimation was based on fitting the isochrones from three models: Yonsei-Yale (YY), PARSEC \citep[the PAdova \& TRieste Stellar Evolution Code,][]{bre12}, and Dartmouth \citep{dot07} to the data from both solutions (A and B). Isochrones calculated for $[M/H]$ = 0.35 presented on three planes ($M_\mathrm{bol}$ - mass, log $T_\mathrm{eff}$ - mass, log $g$ - mass) and shown in Fig. \ref{184949_iso_kompil} favour the solution B (with $T_\mathrm{eff1}$ = 6630 K) and imply a system age of 1~Gyr. 

\begin{figure}
\begin{center}
\includegraphics[angle=-90, width=0.5\textwidth,clip=true]{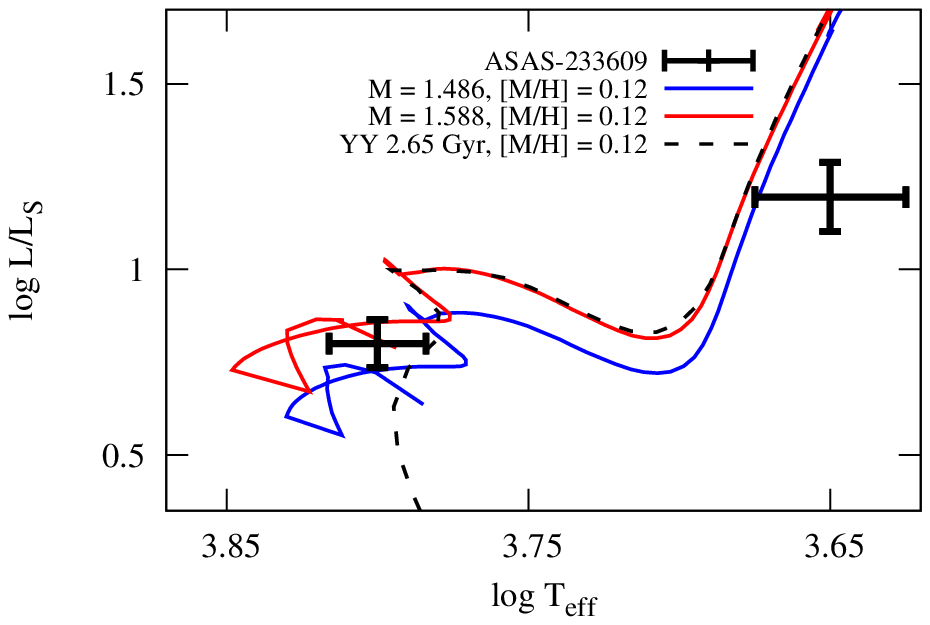}
\includegraphics[angle=-90, width=0.5\textwidth,clip=true]{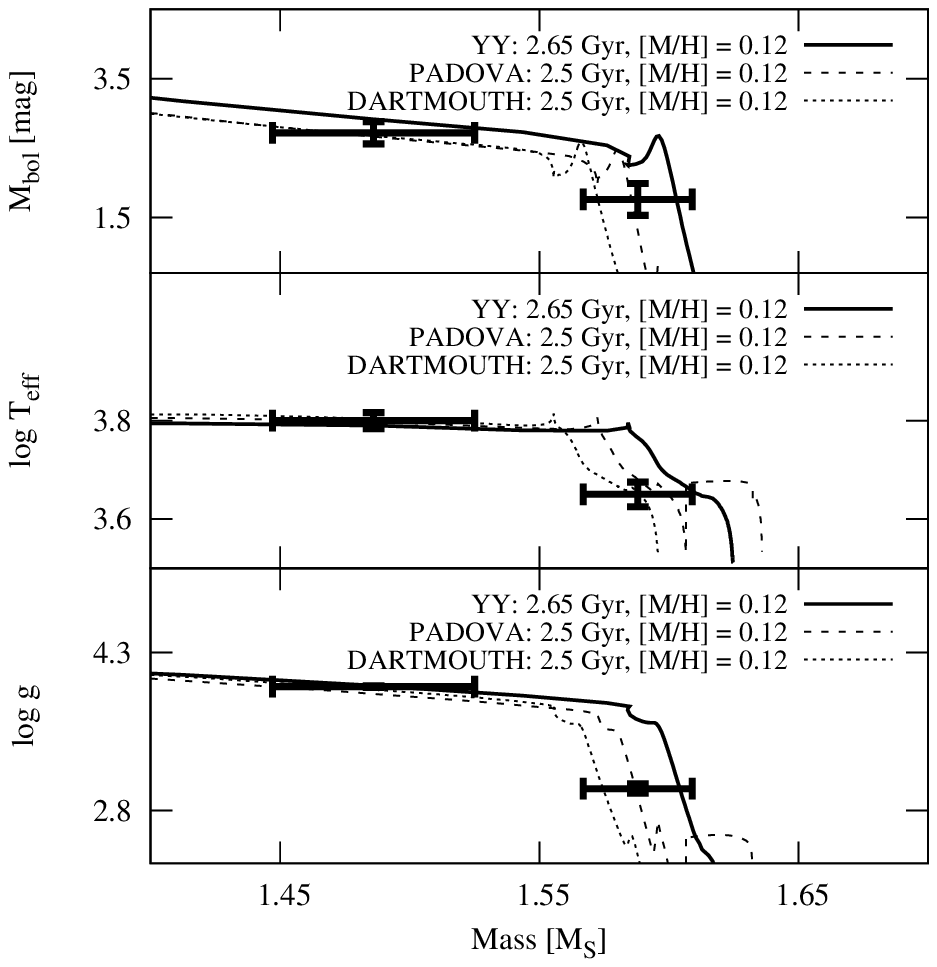}
\end{center}
\caption{YY evolutionary tracks for BQ Aqr and $[M/H]$ = 0.12 with corresponding isochrone (upper panel), and YY, PARSEC, and Dartmouth isochrones of age of 2.5--2.65 Gyr and $[M/H]$ = 0.12 (lower panel).} 
\label{233609_track_iso}
\end{figure}

Evolutionary tracks calculated for the masses of the BQ Aqr components and the system metallicity $[M/H]$ = 0.12 (presented in the upper panel of Fig. \ref{233609_track_iso}) also imply that the components of the system have different evolutionary phases -- the primary is a main sequence star, while the secondary is a red giant. YY, PARSEC, and Dartmouth isochrones (right panel of Fig. \ref{233609_track_iso}) interpolated for $[M/H]$ = 0.12 and ages of 2.65, 2.5, and 2.5 Gyr respectively fit the system parameters, yielding an estimated system age of 2.5--2.65 Gyr. Values of $L$ and $M_\mathrm{bol}$ presented in Fig. \ref{233609_track_iso} were obtained by assuming the system metallicity as $[M/H]$ = 0.12 and $T_1$ = 6390 K. 

\begin{figure*}
\begin{tabular}{cc}
\includegraphics[angle=-90, width=0.5\textwidth,clip=true]{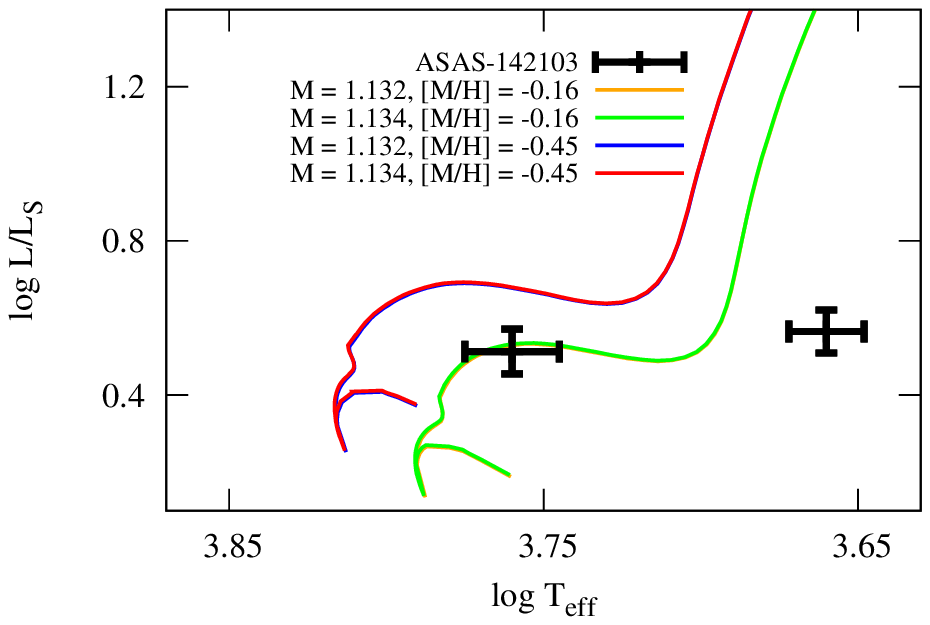}
\includegraphics[angle=-90, width=0.5\textwidth,clip=true]{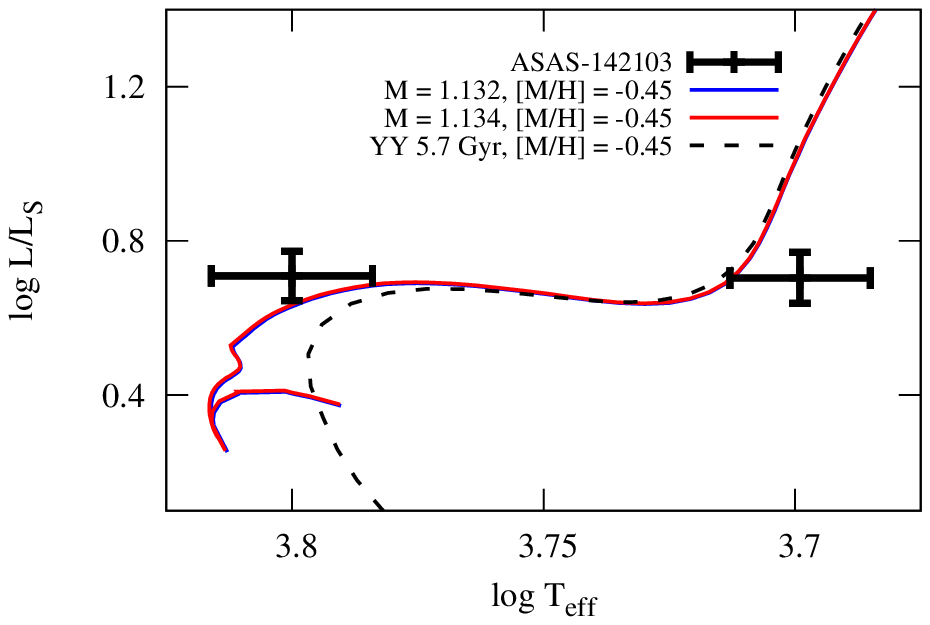}
\end{tabular}
\caption{YY evolutionary tracks for V1207 Cen for the initial solution (left panel) and the alternative one (right panel).}
\label{142103_track}
\end{figure*}

\begin{figure}
\includegraphics[angle=-90,width=0.99\columnwidth]{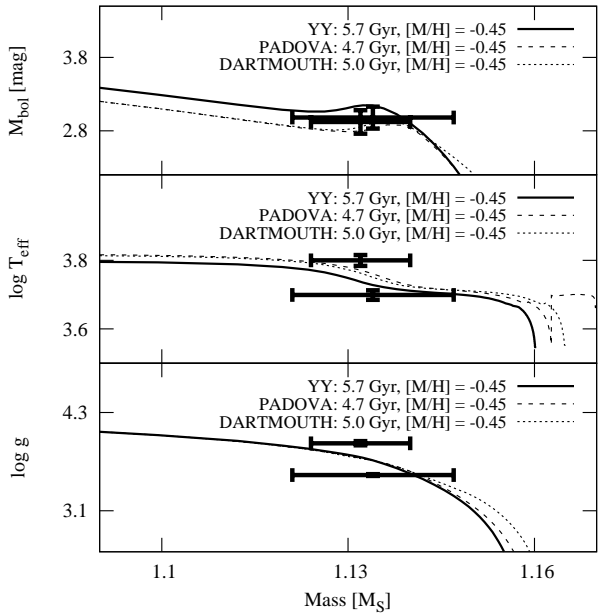}
\caption{YY, PARSEC, and Dartmouth isochrones for $[M/H]$ of -0.45 and age of 5.7, 4.7, and 5.0 Gyr compared with V1207 Cen parameters solution.}
\label{142103_iso}
\end{figure}

Comparison of our results for V1207 Cen with YY isochones led to another inconsistency. As the \textsc{sme} metallicity estimation was not consistent for both components ($[M/H]_{1}$ = -0.16, $[M/H]_{2}$ = -0.45), we checked if our solution fitted evolutionary tracks calculated for both values of metallicity (left panel of Fig. \ref{142103_track}). Any pair of tracks fitted the parameters for both components equally well, offering no solution to the degeneracy in metallicity. We therefore recomputed the effective temperatures by fixing both values of metallicity (as per solution B for ASAS-184949), unfortunately with no improvement of models versus observations. In the next step, we checked if tracks calculated for a wide spread of metallicities fitted the stellar parameters. As we failed in observations and model comparison, we concluded the secondary temperature should be higher. By stepwise tweaking the secondary temperature in \textsc{phoebe} and keeping it fixed, we fitted the value of the primary one, and recalculated $L$ and $M_\mathrm{bol}$ for both components. Such a solution was then compared with evolutionary tracks calculated for either values of metallicity $[M/H]_{1}$ = -0.16 and $[M/H]_{2}$ = -0.45. After a few iterations we found a solution consistent with the models: $T_1$ = 6340 and $T_2$ = 5000 K and the metallicity of $[M/H]$ = -0.45. As shown in Fig. \ref{142103_track} (right panel), evolutionary tracks match the redetermined parameters. However, the intersections of the tracks and the isochrone (obtained for a given metallicity and age determined from comparison presented in Fig. \ref{142103_iso}) indicates that the stellar temperatures should be similar, which is in disagreement with LC solution and unequal depths of eclipses.  This discrepancy may be caused e.g. by a slightly different value of the mass ratio (within the uncertainty from RV solution presented in Table 2) than the one we obtained from our RV solution or even by the mass loss of the more evolved (secondary) component. In such a case the earlier assumption of negligible mass loss could have been invalid. Stellar parameters calculated for the new values of effective temperature are presented in Tab. \ref{tab_orb} as an alternative solution after the slash. The location of the redetermined parameters on evolutionary tracks indicates differential evolutionary phases of the system components -- the primary star is a subgiant, while the secondary is already at the low red giant branch. 

We also compared our redetermined results with YY, PARSEC, and Dartmouth isochrones for $[M/H]$ = -0.45 which yielded system ages of $\sim$ 5.7 Gyr, 4.7 Gyr, and 5.0~Gyr, respectively. The mismatch between the location of the primary star parameters on the plane of log $g$ - mass in Fig. \ref{142103_iso} suggests the value of log $g$ should be lower.

\subsection{Giants with and without spots in eclipsing binaries}

\begin{figure}
\includegraphics[angle=-90,width=0.99\columnwidth]{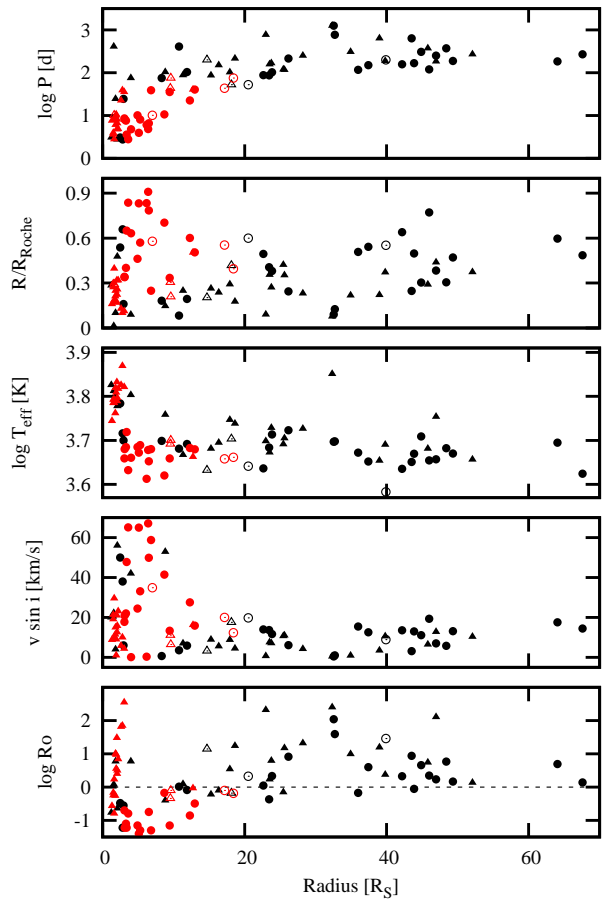}
\caption{Comparison of eclipsing binaries with giants that also show strong
spot-originated brightness modulation (red), with these that do not show it (black).
Circles mark larger (usually cooler) components, and triangles the smaller (hotter) ones.
Open symbols mark our unpublished systems. Orbital period $P$, fractional Roche 
lobe radius $R/R_\mathrm{Roche}$, effective temperature $\log(T)$, projected rotational 
velocity $v \sin i$, and Rossby number $Ro$ are plotted as a function of the 
absolute radius. Note there are no heavily-spotted giants larger than $\sim$20~R$_\odot$ 
and with $Ro>1$ (marked with the dashed line).}\label{fig_spot_comp}
\end{figure}

In Fig. \ref{fig_spot_comp} we compare those eclipsing binaries with giant and subgiant 
components that show spots, with those that do not. Such comparisons in the literature are usually done for main sequence stars, not giants, and using other activity
indicators, like the $L_X/L_{bol}$ or the calcium emission flux ($R_{HK}$). Numerous
studies \cite[e.g.][]{noy84,ste94,gun98,piz03,rei14} relate these 
indicators to rotation period, as the fast rotation drives the dynamo mechanism,
responsible for building strong magnetic fields, or to the Rossby number $Ro$ -- the ratio of the
rotation period to the turnover time for the bottom layer convection zone $\tau_c$
\citep{noy84}. Similar studies that include more evolved stars are sparse 
\citep[e.g.][]{hal94,gun98,gon07}. Much faster evolution after the main sequence phase causes rapid variations in the internal structure, therefore the description of activity is more complicated than for dwarfs \citep{ste94}.

For our compilation, we selected systems from our other works, 
the on-line catalogue DEBCat \citep{sou14}, and also eclipsing
binaries with giant and subgiant components ($R>3$~R$_\odot$) from the Catalogue of
Chromospherically Active Binary Stars \citep[CCABS;][]{eke08}, for which reliable 
fundamental parameters can be found in the literature. Our data is not sufficient 
for studying $L_X/L_{bol}$ or $R_{HK}$, therefore we use a simple photometric indicator of 
`spottedness', similar to \citet{hal94} or \citet{har09}. As `showing spots' we 
define these systems for which the spot-originated modulation is visible in ASAS light
curves.  Because those data are not of the best possible quality, the modulation's 
amplitude is typically $\ga$0.1~mag in $V$-band. If there is no ASAS light curve available, 
we check the literature and look for brightness variations of such a scale. 
The `non-spotted' systems we have chosen typically have photometry of 
better quality than ASAS, and out-of-eclipse brightness fluctuations 
are not clearly seen. 

The `spotted' sample is built of the three targets described here, V1980~Sgr and
ASAS~J010538-8003.7 from \citet{rat13}\footnote{ASAS-010538 was originally
considered as `non-spotted', but a closer inspection of the ASAS light curve 
revealed spots that clearly evolve in time.}, and a number of so-called `classical
RS~CVn-type stars', namely: RZ~Eri, GK~Hya, CF~Tuc, RU~Cnc, VV~Mon, SS~Cam, CQ~Aur, 
SS~Boo, V792~Her, SZ~Psc, LX~Per, and RS~CVn itself\footnote{This list is incomplete, but we consider it as representative, because other RS~CVn-type 
stars share similar characteristics, and the conclusions we present later would 
not change.}. Their physical and orbital parameters were taken from compilations made by 
\citet{pop88,pop90}, and supplemented by works of other authors 
\citep{imb78,arn79,cer80,tum85,fek91,nel91,bur92,eat93,you93,hec95,kan03}, with the exception of CF Tuc, which analysis was revised by \citet{dog09}. We also added to the sample three unpublished systems we have analysed (ASAS-06, ASAS-11, ASAS-16), one of which does not have reliable temperature estimations (ASAS-16).

The `non-spotted' sample contains
14 LMC and SMC systems \citep{gra12,pie13,pil13,gra14}, KIC~8410673 \citep{fra13}, 
AI~Phe \citep{and88,hel09}, TZ~For \citep{and91}, OW~Gem \citep{gal08}, $\alpha$~Aur
\citep{tor09,tor15}, CF~Tau \citep{lac12}, V432~Aur \citep{siv04}, HD187669 
\citep{hel15}, ASAS~J180057-2333.8, \citep{suc15}, ASAS~J182510-2435.5 \citep{rat13},
and two more systems we have analysed but not published yet (ASAS-061, V64).
We want to note, that the `non-spotted' systems may also be active, but the 
activity level must be lower than for the `spotted' ones, as there is no clear 
sign of variability in the published photometry, usually more precise than the ASAS data.
Some of them appear in the CCABS.

In Fig. \ref{fig_spot_comp} we plot the orbital period, ratio of the stellar radius to the Roche radius, effective temperature, projected velocity of rotation, and Rossby number as a function of 
absolute radius. The Roche radius is defined as in Sec. \ref{res}. 
Rotation velocities, $v\sin i$, were either taken directly from the literature 
(for the non-eclipsing $\alpha$~Aur, it was calculated directly from the
rotation periods and radii), or calculated assuming 
(pseudo-)synchronisation\footnote{Pseudo-synchronisation is a state of an 
equilibrium between rotation and orbital motion, achieved for non-circular 
orbits.}, which was explicitly stated in some sources. In very few cases, 
when no information on rotation was found, synchronisation
was also assumed (except the classical cepheid in OGLE-LMC-CEP-0227,
which does not appear in Figure~\ref{fig_spot_comp}). Rossby number $Ro$
values were estimated using the stellar models of \citet{leg12} and $\tau_c$ values from 
grids calculated by the group of C. Charbonnel (private communication).

Figure \ref{fig_spot_comp} shows that the `spotted' giants reside only in
relatively short-period systems, and are all smaller than 20~R$_\odot$. Through 
the assumption of tidal synchronisation, both these facts translate into low
Rossby numbers ($<$1) for the spotted stars, but there are a few cases of 
large ($R>20$~R$_\odot$) stars with equally low $Ro$. The vast majority of the
spotted ones, that clusters
at $R < 10$~R$_\odot$ and $P<10$~d, are the classical RS CVn-type systems.
We suspect that for our objects, despite what the 
best-fitting solutions suggest, only the larger components truly show spots, 
as they are cooler than $\sim$5000~K, and rotating faster than smaller components (except for 
V1980~Sgr, composed of two very similar stars). They also occupy large parts 
of their Roche lobes. However, many `non-spotted' systems show similar values of 
$R/R_\mathrm{Roche}$, $T_\mathrm{eff}$ and $v\sin i$, but are 
significantly larger and on long-period orbits, thus their Rossby numbers are 
usually higher. There are no spotted subgiants and giants (more evolved components of the systems) larger than 20~R$_\odot$ and with $Ro>1$.
The $v\sin i$ alone can not explain the existence or absence of spots,
as some of the smaller spotted stars rotate with velocities between 10
and 20~km~s$^{-1}$, as do many of the larger non-spotted objects.

The explanation for the presence or absence of spots (high and low activity) 
is relatively well understood, and was proposed several decades ago \citep{pop77}. 
A star cools down after the main sequence, expands and becomes tidally locked, 
which increases the velocity of rotation. As the temperature drops, the convective 
envelope expands. A dynamo-like mechanism causes an increase in the magnetic 
field and activity, which is manifested by spots and emission lines. This is a widely-accepted scenario of the formation of active, RS CVn-type
binaries \citep{gou13}. This is also somewhat similar to the situation
of late-type dwarfs in short-period eclipsing binaries, which trapped 
in tidal locking, show high activity levels, and which usually exhibit significant
discrepancies between observed and predicted radii and temperatures \citep{rib08}.

As shown by \citet{hal91,hal94}, larger, spot-originated brightness variability 
of dwarfs is correlated with lower values of $Ro$. He found a threshold at $R\simeq0.65$
between photometrically stable and variable stars. A similar situation can
be seen in Fig. \ref{fig_spot_comp}, where we arbitrarily set the threshold at
$Ro\simeq1$ ($P_{rot}\simeq\tau_c$), but only for stars smaller than 
20~R$_\odot$.
%Stars from our sample that are larger than 20~R$_\odot$ can only be found in long-period binaries, and with higher Rossby numbers, due to their slow (long-period) rotation. 

We conclude that the activity in giants and sub-giants seems to be working in a 
similar way as in dwarfs, where it is most likely driven by some form of dynamo 
mechanism \citep{fei12,fei13}.
We can thus suspect that highly-active giants may exhibit similar discrepancies to
the dwarfs when compared to evolutionary models, i.e. oversized radii and lower
effective temperatures \citep{rib08,hel11}.
The difference is that giants larger than 20~R$_\odot$ tend to be less active,
despite having other parameters ($v\sin i, Ro$) which are similar to those for the smaller and active ones,
suggesting that the size or internal structure plays role in sustaining or suppressing
the activity.
On the other hand, as pointed out by \citet{ste94}, giants' 
activity may be more complicated, taking into account relatively rapid changes in 
their internal structure, but the cut-off in radii seems to be real. 
It is important to emphasize the criterion we have used to distinguish between active (spotted) and less-active (non-spotted) stars is not very robust, and stars we marked as ‘non-spotted’ sometimes show other signs of activity. For a more complete picture of red giants’ activity, additional active, evolved systems should be identified and studied. Optimally, cases of spotted and non-spotted, short-period, 
large and very large ($>$20~R$_\odot$) giants should be observed, to investigate if 
high $Ro$ and $R$ are the only factors that suppress the activity in evolved stars. 
Further observations, and measurements of the intensity of magnetic fields should also be done, and would probably help to determine which kind of dynamo mechanism works in giant, and presumably 
in dwarf stars.

\section{Conclusions}
\label{sec_con}

RV curves and full sets of orbital and physical parameters for three DEB systems (ASAS-184949, BQ Aqr, and V1207 Cen) are presented for the first time. For BQ Aqr and  V1207 Cen we obtained masses and radii values with uncertainties of 1-2 and 2-4 per cent, respectively. Such precision could make these systems useful test beds for empirical verification of stellar evolution models. Despite this high precision, finding a consistent solution for V1207 Cen proved non-trivial. Due to the lack of WASP photometry, the radii in ASAS-184949 are determined to only 20--40 per cent, but mass errors are of $\sim$ 2--4 per cent.

The efficient phase coverage of the RVs made it possible to apply spectral disentangling and spectral analysis yielding independently effective temperatures, and metallicities of the systems' components. As we found inconsistent metallicities for the two components, we decided not to perform an analysis of the  chemical abundances. 

All three systems we analysed consist of a red giant and a less evolved star (in a main sequence or a subgiant phase). Well-characterized detached systems which have not undergone mass transfer, and which consist of components in a slightly different degree of evolution are not very common \citep{sou14}. Fitting models of the same age to the observed properties of both components provides a very stringent test of the models, therefore all of the systems constitute very informative objects worthy of more detailed analysis.

Chromospheric activity of the systems' components, manifested in the existence of spots, is significant for every single system we analysed. Activity is clearly visible in the spectra of A184949 and V1207 Cen, as well as in the shape of the cross correlation function of BQ Aqr, and causes time-varying LC variations in every system. Applying models with spots reproduced the photometric measurements with great performance. Taking into account the stellar parameters and activity of the analysed objects, they can be also classified as classical RS CVn stars \citep{hal76}.

By comparing the properties of giants with and without spots, we found a hint of an activity cut-off at $Ro\sim1$, and $R\sim20$~R$_\odot$. Nevertheless, the details of processes responsible for induction and suppression of stellar activity are still an open question. If the activity observed in the studied systems is induced by rotation and tidal locking, as in dwarfs, further observations of new active, evolved systems might help to distinguish which dynamo mechanism is responsible for the enhanced stellar activity, and resolve the discrepancies between observed and theoretically predicted fundamental stellar parameters.

During the analysis we found several issues troublesome. Metallicities derived from spectral analysis are inconsistent between the two components of ASAS-184949 and do not agree with evolutionary tracks for given masses. Also, the temperature determination is particularly unsatisfactory (solution A for ASAS-184949, V1207 Cen). The obvious complication in determining the temperature are evolving spots, which change the stellar temperature as the star rotates. Spots also affect the outcome of spectral disentangling, as a final decomposed spectrum of a single star is a product of averaging spectra taken with spots visible from different angles.

Another reason for the inconsistencies we met is the brightness ratio, which -- due to the spots' existence -- is not trivial to determine from LC even for systems with total eclipses (spots cause changes in eclipse depths, so it is not easy to determine the exact decrease of total brightness). Thus in the modelling we performed, we took the $BR$ value from the TODCOR analysis (which was consistent with the LC-based $BR$ for the base season without spots). This value was used to scale the disentangled spectra, therefore the entire spectral analysis was based on it.

The analyses we carried out for BQ Aqr and V1207 Cen enrich a relatively small sample of well-characterized red giants, and systems with components in slightly different phases of evolution. In order to improve the radii determination for ASAS-184949, more precise photometry is required. Multicolour photometry for all three systems would be useful to study evolving spots and improve stellar temperature estimation. Infrared observations (both photometric and spectroscopic) could both enrich the study of spots, and allow the investigation of a spectral region where more chromospheric activity indicators are hidden.

\section*{Acknowledgments}

The authors would like to thank M. Armano from ESA for valuable comments, the group of C. Charbonnel from Geneva Observatory for sharing their grids, and the referee for constructive suggestions which helped to improve the manuscript.

This work is supported by the National Science Center through grants 2011/01/N/ST9/02209 (MR), 5813/B/H03/2011/40 (MK), and 2011/03/B/ST9/01819 (KGH), by the European Research Council through a Starting Grant, and by the Foundation for Polish Science through "Idee dla Polski" funding scheme. KGH acknowledges support provided by the National Observatory of Japan as Subaru Astronomical Research Fellow.

This research has made use of the Simbad database, operated at CDS, Strasbourg, France. 

The authors wish to recognize and acknowledge the very significant cultural role and reverence that the summit of Mauna Kea has always had within the indigenous Hawaiian community. We are most fortunate to have the opportunity to conduct observations from this mountain.

\appendix

\section{RV measurements for ASAS-184949, BQ Aqr, and V1207 Cen systems}

The section includes Tables \ref{RV_table_184949}--\ref{RV_table_142103} with RV measurements, formal RV errors, O-Cs, exposure times for each spectrum, SNR and telescope specifications for both components of the selected systems. The used telescopes/spectrographs are as follows: CTIO/CH = CTIO 1.5-m/CHIRON ( f -- fibre mode, s -- slicer mode) EUL/C = Euler/CORALIE, SUB/HDS = Subaru/HDS (red or blue CCD chip), ESO/F = MPG/ESO 2.2-m/FEROS. SNR stands for a signal-to-noise ratio per collapsed spectral pixel at $\lambda$=5\,500~\AA, except HDS red chip ($\lambda$=6\,070~\AA) and HDS blue chip ($\lambda$=4\,760~\AA)  measurements. 

\begin{table*}
\caption{RV measurements for ASAS-184949.}
\centering
\begin{tabular}{c c c c c c c c c c}
\hline \hline
BJD-2450000 & $RV_{1}$ & $\sigma_\mathrm{RV_{1}}$ & $O-C_{1}$ & $RV_{2}$ & $\sigma_\mathrm{RV_{2}}$ & $O-C_{2}$ & $T_\mathrm{exp}$ & SNR & Tel./Sp.\\
 & [km~s$^{-1}$] & [km~s$^{-1}$] & [km~s$^{-1}$] & [km~s$^{-1}$] & [km~s$^{-1}$] & [km~s$^{-1}$] & [s]\\
\hline
 5778.82120 & -7.5 & 1.3 & 3.1 & 26.5 & 1.4 & -0.8 & 600 & 71 & SUB/HDS red \\
 5778.82120 & -9.2 & 1.8 & 1.4 & 27.6 & 2.4 & 0.3 & 600 & 55 & SUB/HDS blue\\
 5855.70366 & -46.6 & 0.8 & -0.9 & 59.0 & 1.6 & 1.0 & 600 & 110 & SUB/HDS red \\
 5855.70366 & -45.9 & 0.6 & 0.3 & 58.1 & 2.3 & 0.1 & 600 & 100 & SUB/HDS blue\\
 6084.67520	& 42.6 & 1.5 & 0.3 & -24.6 & 11.0 & 2.0 & 660 & 28 & EUL/C\\
 6085.69223	& 49.2 & 1.1 & 0.3 & -34.4 & 7.0 & -2.0 & 660 & 32 & EUL/C\\
 6102.72019 & -31.8 & 0.5 & 0.4 & 42.2 & 1.3 & 1.6 & 600 & 80 & ESO/F\\
 6178.16692 & -48.9 & 3.0 & 0.7 & 55.0 & 3.0 & 1.2 & 720 & 10 & EUL/C\\
 6179.09118 & -48.5 & 1.4 & 1.2 & 56.1 & 4.0 & 2.2 & 660 & 25 & EUL/C\\
 6195.08229 & 60.1 & 0.8 & 0.6 & -45.6 & 9.0 & -6.0 & 720 & 90 & ESO/F\\
 6397.75597 & -30.8 & 1.6 & 0.4 &  36.5 & 8.0 & -1.1 & 900 & 31 & EUL/C\\
 6398.83712 & -21.7 & 1.3 & 0.4 & 28.3 & 1.9 & -1.9 & 900 & 28 & EUL/C\\
 6428.69235 & -49.1 & 0.8 & -0.3 & 56.5 & 3.0 & 1.4 & 600 & 53 & ESO/F\\ 
 6497.78145 & -47.5 & 0.9 & -1.9 & 48.7 & 2.4 & -1.4 & 900 & 24 & EUL/C\\
 6498.63918 & -47.7 & 1.1 & 0.4 & 54.8 & 1.9 & 2.3 & 900 & 29 & EUL/C\\
 6506.77702 & -13.2  & 1.4 & 2.2 & 25.9 & 1.4 &  -1.6 & 700 & 50 & CTIO/CH f\\
 6512.67112 & 38.7 & 1.1 &  0.0 & -21.3 & 3.0 & -1.6 & 700 & 50 & CTIO/CH f\\
 6515.71922 &	56.7 & 0.7 & 0.2 & -36.4 & 2.7 & -1.1 & 700 & 50 & CTIO/CH f\\
 6517.66739 & 60.5 & 0.9 & 0.0 & -40.3 & 6.0 & -1.5 & 700 & 50 & CTIO/CH f\\
 6517.68720 & 60.4 & 1.0 & -0.8 & -43.3 & 8.0 & -2.1 & 720 & 41 & ESO/F\\
 6519.71238 & 57.8 & 0.9 & -0.7 & -42.8 & 4.0 & -4.0 & 600 & 63 & ESO/F\\
 6519.72110 & 57.8 & 0.8 & -0.7 & -42.4 & 2.2 & -4.0 & 500 & 55 & ESO/F\\
 6522.65524 & 42.6 & 0.9 & 0.3 & -21.4 & 0.6 & 1.6 & 700 & 50 & CTIO/CH f\\
 6532.64753 & -41.5 & 1.9 & -0.1 & 49.0 & 1.4 &  -1.4 & 700 & 50 & CTIO/CH f\\
 6547.53627 &  31.4  & 0.8 & -0.5 & -14.6 & 2.2 & -0.9 & 700 & 50 & CTIO/CH f\\
 6554.53677 & 60.1 & 1.0 & 0.3 & -42.7 & 8.0 & -4.5 & 700 & 50 & CTIO/CH f\\
 6555.53447 & 57.7 & 0.8 & -0.5 & -41.0 & 2.0 & -4.9 & 700 & 50 & CTIO/CH f\\
 6577.51728 & -21.0 & 1.7 & 0.3 & 30.2 & 2.2 & -2.6 & 700 & 50 & CTIO/CH f\\
 6582.50199 & 22.7 & 2.5 & -2.6 & -8.6 & 2.5 & 0.6 & 700 & 50 & CTIO/CH f\\
 6582.53138 & 21.8 & 1.7 & -3.9 & -10.1 & 3.0 & -1.8 & 700 & 50 & CTIO/CH f\\
 6584.51811 & 41.2 & 2.4 & -0.8 & -24.8 & 3.0 & -2.2 & 700 & 50 & CTIO/CH f\\ 
 
\hline
\label{RV_table_184949}
\end{tabular}
\end{table*}

\begin{table*}
\caption{RV measurements for BQ Aqr.}
\centering
\begin{tabular}{c c c c c c c c c c}
\hline \hline
BJD-2450000 & $RV_{1}$ & $\sigma_\mathrm{RV_{1}}$ & $O-C_{1}$ & $RV_{2}$ & $\sigma_\mathrm{RV_{2}}$ & $O-C_{2}$ & $T_\mathrm{exp}$ & SNR & Tel./Sp.\\
 & [km~s$^{-1}$] & [km~s$^{-1}$] & [km~s$^{-1}$] & [km~s$^{-1}$] & [km~s$^{-1}$] & [km~s$^{-1}$] & [s]\\
\hline 
 5846.71804  & -53.6  &  0.8 &  0.0  & 86.3 &  9.0  & -2.4  & 750 & 20 & EUL/C\\
 5893.55044  & -26.1  &  0.5  &  0.1 &  65.9 & 6.0 &  3.6 & 960 & 13 & CTIO/CH s\\
 6079.92999  & 52.3  & 0.3  & 0.3  & -6.9  & 9.0   & 4.0 & 1270 & 20 & CTIO/CH s\\
 6080.92988  & 101.9  &  1.4 & -0.3 & -50.9 &  10.0  & 6.2 & 750 & 13 & EUL/C\\
 6081.86515  & 86.1  &  0.6  & -0.9  & -34.1  &  6.0  &  8.1  & 750 & 23 & EUL/C\\
 6083.93994  & -56.2 & 0.4  & -0.7  & 86.3  &  6.0 &  -3.4 & 1270 & 26 & CTIO/CH s\\
 6084.88699  & -59.2 &  0.3 & 0.4  & 95.9  & 6.0 & 1.6  & 900 & 24 & EUL/C\\
 6179.72531  & 84.2  &  0.4  &  0.3 & -40.7  &  5.0  & -0.7  & 900 & 25 & EUL/C\\
 6180.78603  & 99.9 &  0.4 & -0.6  & -50.9  &  7.0  &  4.7  & 900 & 28 & EUL/C\\
 6192.74475  & 70.9  & 0.3  & 0.1  & -29.3  & 3.0   & -0.9  & 1200 & 27 & CTIO/CH s\\ 
 6242.64302  & -47.6 &  0.6  & -0.7  & 83.8  &  6.0  &  1.4  & 900 & 30 & EUL/C\\
 6497.72638  & 94.1	& 0.6 & 0.1 & -54.2 & 6.0 & -4.7  & 900 & 22 & EUL/C\\
 6497.89228 & 99.4	& 0.3 & -0.1 & -57.3 & 3.0 & -2.8 & 900 & 31 & EUL/C\\
 6498.84591 &  91.2	& 0.4 & 0.2 & -41.6	& 12.0 & 5.0 & 900 & 24 & EUL/C\\ 
 6506.89276  & -10.4  &  0.3 & -0.1  & 39.7  &  3.0  & -6.3 & 750 & 45 & CTIO/CH f\\ 
 6508.93344  &-45.5  &  0.2 & -0.0  & 85.1  & 5.0  &  6.3  & 750 & 45 & CTIO/CH f\\
 6512.72126  & 51.9  & 0.4  & -0.2  & -9.8  & 5.0 &  2.7  & 750 & 45 & CTIO/CH f\\
 6514.84285  &-66.2 &  0.3 & 0.07  & 95.4  &  3.0  & -2.8 & 750 & 45 & CTIO/CH f\\ 
 6538.75587  & 82.0  & 0.3  & 0.6 & -40.4  &  8.0  & -0.4 & 750 & 40 & CTIO/CH f\\ 
 6547.84414  &-66.3 &  0.3  & -0.1  & 92.0  &  4.0  & -6.2 & 750 & 40 & CTIO/CH f\\ 
 6550.65153  & 91.9  & 0.4  & -0.4 & -52.8  &  3.0  & -2.8 & 750 & 35 & CTIO/CH f\\ 
 6556.54973  & 47.4  & 0.4  & -0.9  & -5.4  &  3.0  &  3.6  & 750 & 45 & CTIO/CH f\\
 6568.51741  &-44.9 &  0.4  & -0.9  & 88.4  &  5.0   & 9.7  & 750 & 40 & CTIO/CH f\\
 6570.61228  & 95.9  & 0.3  & -0.1  & -52.4  &  3.0  & -2.7 & 750 & 40 & CTIO/CH f\\ 
 6575.52576  & -20.6 & 0.8  &  0.4  & 45.5  & 6.0  & -10.4 & 750 & 30 & CTIO/CH f\\ 
 6581.52829  & -56.0  & 0.5  & -0.1 &  93.6  &  2.8 &   4.9  & 750 & 35 & CTIO/CH f\\
 6583.53373  & 81.5  & 0.5 & -0.2 & -43.9  & 6.0 &  -3.8 & 750 & 35 & CTIO/CH f\\ 
\hline
\label{RV_table_233609}
\end{tabular}
\end{table*}

\begin{table*}
\caption{RV measurements for V1207 Cen.}
\centering
\begin{tabular}{c c c c c c c c c c}
\hline \hline
BJD-2450000 & $RV_{1}$ & $\sigma_\mathrm{RV_{1}}$ & $O-C_{1}$ & $RV_{2}$ & $\sigma_\mathrm{RV_{2}}$ & $O-C_{2}$ & $T_\mathrm{exp}$ & SNR & Tel./Sp.\\
 & [km~s$^{-1}$] & [km~s$^{-1}$] & [km~s$^{-1}$] & [km~s$^{-1}$] & [km~s$^{-1}$] & [km~s$^{-1}$] & [s]\\
\hline
 6692.88109 & -6.3 & 0.2 & 0.0 & -142.6 & 0.8 & -0.7 & 900 & 50 & CTIO/CH f\\ 
 6736.65885 & -26.9 & 0.2 & 0.3 & -120.4 & 1.0	& 0.6 & 900 & 50 & CTIO/CH f\\ 
 6749.69003 & -113.1 & 0.2 & 0.1 & -33.9 & 0.8 &	0.9 & 900 & 50 & CTIO/CH f\\ 
 6769.67348 & -6.4 & 0.2 & -0.0 & -140.8 & 0.8 & 1.1 & 900 & 50 & CTIO/CH f\\ 
 6790.56491 & -138.8 & 0.3 & -0.3 & -9.8 & 0.7 & -0.3 & 900 & 50 & CTIO/CH f\\ 
 6840.58050 & -96.2 & 0.3 & 0.0 &  -51.7 & 0.7	& 0.1 & 900 & 50 & CTIO/CH f\\ 
 6869.52784 & -98.4 & 0.7 & 0.6 &  -50.7 & 1.8	& -1.6 & 900 & 50 & CTIO/CH f\\ 
 6898.49168 & -15.9 & 0.2 & -0.2 & -133.1 & 1.3 	& -0.6 & 1000 & 60 & CTIO/CH f\\ 
 6904.49056 & -58.8 & 0.1 & -0.1 &   -90.2 & 0.6 &	-0.7 & 1000 & 60 & CTIO/CH f\\
\hline
\label{RV_table_142103}
\end{tabular}
\end{table*}

\bsp

\label{lastpage}


\begin{thebibliography}{99}
\bibitem[Andersen et al.(1988)]{and88}
  Andersen J., Clausen J.~V., Nordstrorm B., Gustafsson B., Vandenberg D.~A., 1988, A\&A, 196, 128 
\bibitem[Andersen et al.(1991)]{and91}
 Andersen J., Clausen J.V., Nordstorm B., Tomkin J., Mayor M., 1991, A\&A, 246, 99
 \bibitem[Arnold et al.(1979)]{arn79} Arnold C. N., Montle R. E., Stuhlinger T. W., Hall D. S., 1979, AcA, 29, 243
\bibitem[Bagnuolo \& Gies(1991)]{bag91}
	 Bagnuolo W.~G., Gies D.~R., 1991, ApJ, 376, 266
\bibitem[Bessell et al.(1998)]{bes98}
	Bessell M.~S., Castelli F., Plez B., 1998, A\&A, 333, 231      
\bibitem[Bressan et al.(2012)]{bre12}	
	Bressan A., Marigo P., Girardi L., Salasnich B., Dal Cero C., Rubele S., Nanni S., 2012, MNRAS, 427, 127 
\bibitem[Burki et al.(1992)]{bur92} Burki G., Kv{\'i}z Z., North P., 1992, A\&A, 256, 463 	
\bibitem[Cerruti-Sola et al.(1980)]{cer80} Cerruti-Sola M. et al., A\&A, 1980, 42, 245%8 aut: Scaltriti, F.; Blanco, C.; Catalano, S.; Marilli, E.; Rodono, M.; Strazzulla, G.; Chambliss, C. R.
%\bibitem[Demarque et al.(2004)]{dem04}
	Demarque P., Woo J.-H., Kim Y.-C., Yi S. K., 2004, ApJS, 155, 667
%\bibitem[di Benedetto(2005)]{diB05}
   di Benedetto G. P., 2005, MNRAS, 357, 174	
\bibitem[Dogru et al.(2009)]{dog09} Dogru D., Erdem A., Dogru S. S., Zola S., 2009, MNRAS, 397, 1647
\bibitem[Dotter et al.(2007)]{dot07}
	Dotter A., Chaboyer B., Jevremovic D., Baron E., Ferguson J. W., Sarajedini A., Anderson J., 2007, AJ, 134, 376	
\bibitem[Eaton et al.(1993)]{eat93} Eaton J. A., Henry G. W., Bell C., Okorogu A., 1993, AJ, 106, 1181
\bibitem[Eggleton(1983)]{egg83} Eggleton P. P., 1983, ApJ, 268, 368
\bibitem[Eker et al.(2008)]{eke08} Eker Z. et al., MNRAS, 2008, 389, 1722%10 aut
%\bibitem[Etzel(1981)]{etz81} 
	Etzel P. B., 1981, Photometric and Spectroscopic Binary Systems, Proc. of NATO Advanced Study Inst. Series, eds. Carling E.B., Kopal Z., publisher: Springer, p. 111
\bibitem[Feiden \& Chaboyer(2012)]{fei12} Feiden G. A., Chaboyer B., 2012, ApJ, 761, 30
\bibitem[Feiden \& Chaboyer(2013)]{fei13} Feiden G. A., Chaboyer B., 2013, ApJ, 779, 183	
\bibitem[Fekel(1991)]{fek91} Fekel F. C., 1991, AJ, 101, 1489
\bibitem[Frandsen et al.(2013)]{fra13} Frandsen S. et al., 2013, A\&A, 556, A138%14 aut
\bibitem[Ga{\l}an et al.(2008)]{gal08} Ga{\l}an C., Miko{\l}ajewski M., Tomov T., Kolev D., Graczyk D., Majcher A., Janowski J.~L., Cikala M.,  2008, Obs., 128, 298  
%\bibitem[Girardi et al(2000)]{gir00}
	Girardi L., Bressan A., Bertelli G., Chiosi C., 2000, A\&AS, 141, 371  
\bibitem[Gould et al.(2013)]{gou13} Gould A. et al., 2013, ApJ, 763, 141%>100 aut
\bibitem[Gondoin(2007)]{gon07} Gondoin P., 2007, A\&A, 464, 1101 
\bibitem[Graczyk et al.(2012)]{gra12}
	Graczyk D. et al., 2012, ApJ, 750, 144	
\bibitem[Graczyk et al.(2014)]{gra14} Graczyk D. et al., 2014, ApJ, 780, 59%15 aut
\bibitem[Grevesse et al.(2007)]{gre07} Grevesse N., Asplund M., Sauval A.~J., 2007, SSRv, 130, 105 
\bibitem[Gunn et al.(1998)]{gun98} Gunn A. G., Mitrou C. K., Doyle J. G., 1998, MNRAS, 296, 150
\bibitem[Hackert \& Ordway(1995)]{hec95} Heckert P. A., Ordway J. I., 1995, AJ, 109, 2169
\bibitem[Hall(1976)]{hal76}
Hall D. S., 1976, ASSL, 60, 287
\bibitem[Hall(1991)]{hal91} Hall D. S., 1991, in: The Sun and Cool Stars: Activity, Magnetism, Dynamos, eds. I. Tuominen, D. Moss, G. R\"udiger, Berlin, Springer Verlag, p. 353
\bibitem[Hall(1994)]{hal94} Hall D. S., 1994, MmSAI, 65, 73
\bibitem[Hartman et al.(2009)] {har09} Hartmann J. D. et al., 2009, ApJ, 691, 342 %9aut
\bibitem[He\l miniak et al.(2009)]{hel09}
	He\l miniak K.~G., Konacki M., Muterspaugh M.~W., Ratajczak M., 2009, 
	MNRAS, 400, 969
\bibitem[He\l miniak \& Konacki(2011)]{hel11}
	He\l miniak K.~G., Konacki M., 2011, A\&A, 526, A29
\bibitem[He\l miniak et al.(2014)]{hel14}
	He\l miniak K.~G., Brahm R., Ratajczak M., Espinoza N., Jordan A., Konacki M., Rabus M., 2014, A\&A, 567, 64		
\bibitem[He\l miniak et al.(2015)]{hel15}
	He\l miniak K.~G. et al., 2015, MNRAS, 448, 1945
%\bibitem[Hoffleit et al.(1962)]{hof62}
    Hoffleit D., 1962, AJ	, 67, 228
\bibitem[Hoffmeister(1931)]{hof31}
	Hoffmeister C., 1931, AN, 242, 129    
\bibitem[H{\o}g et al.(2000)]{hog00}
	H{\o}g E. et al., 2000, A\&A, 355, 27
\bibitem[Ilijic et al.(2004)]{ili04}
	Ilijic S., Hensberge H., Pavlovski K., Freyhammer L., 2004, ASP Conf. Ser. 318, Spectroscopically and Spatially Resolving the Components of Close Binary Stars. Astron. Soc. Pac., San Francisco, p. 111
\bibitem[Imbert(1978)]{imb78} Imbert M., 1978, A\&AS, 33, 321
\bibitem[Jordan et al.(2014)]{jor14}
	Jordan, A. et al. 2014, AJ, 148, 29
\bibitem[Kang et al.(2003)]{kan03} Kang Y. W., Lee W.-B., Kim H.-I., Oh K.-D., 2003, MNRAS, 344, 1227
\bibitem[Kaufer et al.(1999)]{kau99}
	Kaufer, A. et al. 1999, The Messenger, 95, 8	
%\bibitem[Kervella et al.(2004)]{ker04}
  Kervella P., Thevenin F., Di Folco E., Segransan D. 2004, A\&A, 426, 297 	
\bibitem[Klinglesmith \& Sobieski(1970)]{kli70}
	Klinglesmith D.~A., Sobieski S., 1970, AJ, 75, 175
\bibitem[Konacki et al.(2010)]{kon10}
	Konacki M., Muterspaugh M.~W., Kulkarni S.~R., He\l miniak K.~G., 2010, ApJ, 719, 129
\bibitem[Kordopatis et al.(2013)]{kor13}
	Kordopatis G. et al. 2013, AJ, 146, 134	
\bibitem[Kupka et al.(1999)]{kup99}
	Kupka F., Piskunov N., Ryabchikova T.~A., Stempels H.C., Weiss W.~W., 1999, A\&AS, 138, 119
	\bibitem[Kurucz(1992)]{kur92}
	Kurucz R. L., 1992, in IAUSymp., 149, The Stellar Population of Galaxies, ed. B. Barbury, A. Renzini, 225
\bibitem[Lacy et al.(2012)]{lac12} Lacy C. H. S., Torres G., Claret A., 2012, AJ, 144, 167
\bibitem[Lastennet \& Valls-Gabaud(2002)]{las02}
    Lastennet E., Vall-Gabaud D., 2002, A\&A, 396, 551
\bibitem[Legarde et al.(2012)]{leg12} Legarde N., Decressin T., Charbonnel C., Eggenberger P., Ekstrom S, Palacios A., 2012, A\&A, 543, 108
\bibitem[Lucy(1967)]{luc67}
	Lucy L.~B, 1967, Zeitschrift fur Astrophysics, 65, 98	
%\bibitem[Marigo et al.(2008)]{mar08}
	Marigo P., Girardi L., Bressan A., Gronewegen M.~A.~T., Silva L., Granato G. L., 2008, A\&A, 482, 883
%\bibitem[Mayor et al.(2003)]{may03}
	Mayor M. et al., 2003, Messenger, 114, 20         
\bibitem[Nelson et al.(1991)]{nel91} Nelson C. H., Hall D. S., Fekel F. C., Fried R. E., Lines R. E., Lines H. C., 1991, Ap\&SS, 182, 1	
	\bibitem[Noguchi(2002)]{nog02}
	Noguhi, K. et al. 2002, PASJ, 54, 855
\bibitem[Noyes et al.(1984)]{noy84} Noyes R. W., Hartmann L. W., Baliunas S. L., Duncan D. K., Vaughan A. H., 1984, ApJ, 279, 763	
%\bibitem[Paczy\'nski et al.(2006)]{pac06}
	Paczy\'nski B., Szczygie\l~D.~M., Pilecki B., Pojma\'nski G., 2006, MNRAS, 368, 1311    
\bibitem[Penev et al.(2013)]{pen13}
	Penev K. et al., 2013, AJ, 145, 5 	
%\bibitem[Pietrzy\'nski et al.(2009)]{pie09}
	Pietrzy\'nski G. et al. 2009, ApJ, 697, 862 
%\bibitem[Pietrzy\'nski et al.(2010)]{pie10}
 	Pietrzy\'nski G. et al., 2010, Nature, 468, 542 
%\bibitem[Pietrzy\'nski et al.(2011)]{pie11}
	Pietrzy\'nski G. et al., 2011, ApJ, 742L, 20  
\bibitem[Pietrzy\'nski et al.(2013)]{pie13}
	Pietrzy\'nski G. et al., 2013, Nature, 495, 76
\bibitem[Pilecki et al.(2013)]{pil13} Pilecki B. et al., 2013, MNRAS, 436, 953%17 aut
\bibitem[Piskunov et al.(1995)]{pis95}	
	Piskunov N., Kupka F., Ryabchikova T.~A, Weiss W.~W., Jeffery, C.~S., 1995, A \&AS, 112, 525 
\bibitem[Pizzolato et al.(2003)]{piz03} Pizzolato N., Maggio A., Micela G., Sciortino S., Ventura P., 2003, A\&A, 397, 147
\bibitem[Pojma\'nski(2002)]{poj02} 
	Pojma\'nski G. 2002, AcA, 52, 397
\bibitem[Pollacco et al.(2006)]{pollacco}
	Pollacco D.~L. et al., 2006, PASP, 118, 1407
%, Skillen I., Cameron A.~C., Christian D.~J., Hellier C., Irwin J., Lister T.~A., Street R.~A.,  West R.~G., Anderson D., Clarkson W.~I., Deeg H., Enoch B., Evans A., Fitzsimmons A., Haswell C.~A., Hodgkin S., Horne K., Kane S.~R., Keenan F.~P., Maxted P.~F.~L., Norton A.~J., Osborne J., Parley N.~R., Ryans R.~S.~I., Smalley B., Wheatley P.~J., Wilson D.~M., 2006, PASP, 118, 1407
\bibitem[Popper(1988)]{pop88} Popper D. M., 1988, AJ, 96, 1040
\bibitem[Popper(1990)]{pop90} Popper D. M., 1990, AJ, 100, 247
\bibitem[Popper \& Etzel(1981)]{pop81}	Popper D. M., Etzel P. B., 1981, AJ, 86, 102
\bibitem[Popper \& Ulrich(1977)]{pop77} Popper D. M., Ulrich R. K., 1977, ApJ, 212, L131
\bibitem[Pr\v{s}a \& Zwitter(2005)]{prs05} 
	Pr\v{s}a A., Zwitter T., 2005, ApJ, 628, 426
\bibitem[Queloz et al. (2001)]{que01}	
	Queloz D. et al. 2001, The Messenger, 105, 1	
\bibitem[Ratajczak et al.(2013)]{rat13}
	Ratajczak M., He\l miniak K.~G., Konacki M., Jordan A., 2013, MNRAS, 433, 2357
\bibitem[Reiners et al.(2014)]{rei14} Reiners A., Sch{\"u}ssler M., Passegger V. M., 2014, ApJ, 794, 144
\bibitem[Ribas et al.(2008)]{rib08}
	Ribas I., Morales J.~C., Jordi C., Baraffe I., Chabrier G., Gallardo J.,
	2008, MmSAI, 79, 562
\bibitem[Schlafly \& Finkbeiner(2011)]{sch11}
Schlafly E.~F., Finkbeiner D.~P., 2011, ApJ, 737, 103
\bibitem[Schlegel et al.(1998)]{sch98}
 Schlegel D., Finkbeiner D., Davis M., 1998, ApJ, 500, 525
\bibitem[Schroder \& Cuntz(2005)]{sch05}
	Schroder K.~P., Cuntz M., 2005, ApJ, 630, 73 
\bibitem[Schwab et al.(2012)]{sch12}
	Schwab Ch., Spronck J., Tokovinin A., Szymkowiak A., Giguere M., Fisher D., Performance of the CHIRON high-resolution Echelle
spectrograph, 2012, p.16, eds. McLean I. S., Ramsay S. k., Takami H., Proc. SPIE, 8446, paper 8446-9
%\bibitem[Simon \& Sturm(1994)]{sim94}
	Simon K.~P., Sturm E., 1994, A\&A, 281, 286
\bibitem[Siviero et al.(2004)]{siv04} Siviero A., Munari U., Sordo R., Dallaporta S., Marrese P. M., Zwitter T., Milone E. F., 2014, A\&A, 417, 1083
\bibitem[Sitek \& Pojma\'nski(2014)]{sit14}
	Sitek M., Pojma\'nski G., 2014, AcA, 64, 115
\bibitem[Smith et al.(2006)]{smith}
	 Smith A.~M.~S. et al., 2006, MNRAS, 373, 1151
%, Collier Cameron A., Christian D.~J., Clarkson W.~I., Enoch B., Evans A., Haswell C.~A., 	Hellier C., Horne K., Irwin J., Kane S.~R., Lister T.~A., Norton A.~J., Parley N., Pollacco D.~L., Ryans R., Skillen I., Street R.~A., Triaud A.~H.~M.~J., West R.~G., Wheatley P.~J., Wilson D.~M., 2006, MNRAS, 373, 1151
\bibitem[Southworth et al.(2004a)]{sou04a} 
	Southworth J., Maxted P.~F.~L., Smalley B., 2004a, MNRAS, 351, 1277
\bibitem[Southworth et al.(2004b)]{sou04b} 
	Southworth J., Zucker S., Maxted P.~F.~L., Smalley B., 2004b, MNRAS, 355, 986
%\bibitem[Southworth et al.(2005)]{sou05}
	Southworth J., Maxted P.~F.~L., Smalley B., 2005, A\&A, 429, 645	
\bibitem[Southworth (2015)]{sou14}
	Southworth J., 2015, ASPC, 466, 164	
\bibitem[St\c{e}pie{\'n}(1994)]{ste94} St\c{e}pie{\'n} K., 1994, A\&A, 292, 191
\bibitem[Strassmeier et al.(1994)]{str94}
	Strassmeier K.~G., Handler G., Pauzen E., Raith M., 1993, A\&A, 281, 855 
\bibitem[Strohmeier(1966)]{str66}
	Strohmeier, W., 1966, IBVS, 158, 1	
\bibitem[Suchomska et al.(2015)]{suc15} Suchomska K. et al., 2015, MNRAS, 451, 651%14 aut
\bibitem[Tamuz et al.(2005)]{tamuz}
	Tamuz O., Mazeh T., Zucker S., 2005, MNRAS, 356, 1466
\bibitem[Tokovinin et al.(2013)]{tok13}
	Tokovinin et al. 2013, PASP, 125, 1336 
\bibitem[Torres et al.(2006)]{tor06}
	Torres C.~A.~O., Quast G.~R., Da Silva L., De Le Reza R., Melo C.~H.~F., Sterzik M., 2006, A\&A, 460, 695 
\bibitem[Torres et al.(2009)]{tor09} Torres G., Claret A., Young P. A., 2009, ApJ, 700, 1349
\bibitem[Torres et al.(2010)]{tor10}
	Torres G., Andersen, J., Gimen{\'e}z, A., 2010, A\&A Rev, 18, 67
\bibitem[Torres et al.(2015)]{tor15} Torres G., Claret A., Pavlovski K., Dotter A., 2015, ApJ, 807, 26 
\bibitem[T{\"u}mer et al.(1985)]{tum85} T{\"u}mer O., Ibano\v{g}lu C., Evren S., Tunca Z. 1985, Ap\&SS, 112, 273
\bibitem[Valenti \& Piskunov(1996)]{val96}
	Valenti J.~A., Piskunov N., 1996, A\&AS, 118, 595	
\bibitem[Valenti et al.(1998)]{val98}
 	 Valenti J.~A., Piskunov N., Johns-Krull, C.~M., 1998, ApJ, 498, 851 
\bibitem[Valenti \& Fischer(2005)]{val05} 
	Valenti J.~A., Fischer D.~A., 2005, ApJS, 159, 141	
\bibitem[van Hamme(1993)]{van93}
	van Hamme W., 1993, AJ, 106, 2096
\bibitem[Voges et al.(1999)]{vog99}
	Voges W. et al. 1999, A \&A, 349, 389
%\bibitem[Wilson \& Devinney(1971)]{wil71}
	Wilson R.~E., Devinney, E.~J., 1971, ApJ, 166, 605
\bibitem[Worthey \& Lee(2011)]{wor11}
	Worthey G., Lee H., 2011, ApJ, 193, 1 
\bibitem[Yi et al.(2001)]{yi01}
	Yi S.~K., Demarque P., Kim Y.~C., Lee Y.~W., Ree C.~H., Lejeune T., Barnes S.,
	2001, ApJS, 136, 417
\bibitem[Young(1993)]{you93} Young W. K., 1993, Ap\&SS, 201, 35	
\bibitem[Zucker \& Mazeh(1994)]{zuc94}
	Zucker S., Mazeh T., 1994, ApJ, 420, 806


\end{thebibliography}
\end{document}